\documentclass[reprint,superscriptaddress,twocolumn,10pt]{revtex4-1}
\usepackage{bbm}
\usepackage{physics}
\usepackage{amsmath}
\usepackage{epsfig}
\usepackage{subfigure,mathrsfs}
\usepackage{array}
\usepackage{wrapfig}
\usepackage{amssymb}
\usepackage{braket}
\usepackage{float}
\usepackage{lmodern,amssymb}
\usepackage{physics}
\usepackage[dvipsnames]{xcolor}

\newcommand{\beq}[0]{\begin{equation}}
\newcommand{\eeq}[0]{\end{equation}}

\def\be{\begin{equation}}
\def\ee{\end{equation}}
\def\bea{\begin{eqnarray}}
\def\eea{\end{eqnarray}}
\newcommand{\ba}{\begin{eqnarray}}
\newcommand{\ea}{\end{eqnarray}}
\usepackage{hyperref}

\begin{document}
	\title{Universal behaviour of Coulomb coupled Fermionic thermal diode}
	%\author{Shuvadip Ghosh}
	%\author{Nikhil Gupt}
	%\author{Arnab Ghosh}
	%\affiliation{Indian Institute of Technology Kanpur, 
	%Kanpur,Uttar Pradesh 208016, India}
	\author{Shuvadip Ghosh, Nikhil Gupt and Arnab Ghosh \\
		\small \textit{Indian Institute of Technology Kanpur,
			Kanpur, Uttar Pradesh 208016, India}}
	\begin{abstract}
	We propose a minimal model of a Coulomb-coupled fermionic quantum dot thermal diode that can act as an efficient thermal switch and exhibit complete rectification behavior, even in the presence of a small temperature gradient. Using two well-defined dimensionless system parameters, universal characteristics of the optimal heat current conditions are identified. It is shown to be independent of any system parameter and is obtained only at the mean transitions point ``$-0.5$'', associated with the equilibrium distribution of the two fermionic reservoirs, tacitly referred to as ``\textit{universal magic mean}''.	
	\end{abstract}
	
	\maketitle

    \section{Introduction}\label{Sec. I}

	\par Heat management devices have attracted recent interest in nanoscale systems to prevent overheating due to heat flow in desired areas of electronic circuits~\cite{giazotto2006RMP,pekola2015Naturephysics,beneti2017Physicsreport,roberts2011Sciencedirect,li2012RMP,kurizki2020PRR,zeng2008PRB,ruokola2009PRB,kuo2010PRB,liu2014APL,rafael2017NJP,landi2021arxiv}. Following the  theoretical proposal of a quantum dot thermal diode by Roukola and Ojanen~\cite{PRBDIODE2011}, a number of  studies are carried out in achieving rectification effect using simple quantum systems~\cite{terraneo2002PRL,li2004PRL,segak2005PRL,PRB2009,wu2009PRL,PREREC2014,Mascarenhas2014IOP,PRB2018REC,PRB2021,iorio2021PRAPP,balachandran2019PRL,kargi2019PRE,PREREC2017,iorio2021PRAPP,PRB2020,upadhyay2021PRE,diaz_2021NJP,NJP2022}. Yet, most of the works that have been done till now rely on either temperature gradient of bosonic reservoirs or different coupling strengths between the system and the bath to break the inversion symmetry of the overall system~\cite{NJP2022}. For example, Werlang et. al.~\cite{PREREC2014} explored heat transport under the influence of strong coupling between two spins interacting with their respective bosonic bath. Miranda et. al.~\cite{PREREC2017} identified similar diode characteristic in presence of different excitation frequencies between the coupled spins. Based on a two and four terminal quantum dot setups, Tesser et.al.~\cite{NJP2022} recently explored the role of level degeneracy and temperature bias. However, much less attention is paid in achieving rectification effect by means of statistical properties of the reservoir. Here, we provide a general framework to capture the invariant aspect of the fermionic diodes in terms of two dimensionless physical parameters and statistical distribution of the reservoir --- the \textit{uniqueness} of which is found to be independent of the system energy levels, interaction strength, bath spectrum, its temperature and chemical potentials.

	\par To showcase our findings, we consider a Coulomb coupled quantum dot system and study the interaction with two fermionic reservoirs with different temperature and chemical potential. We find while temperature gradient governs the overall heat flow direction, magnitude of heat current is primarily controlled by chemical potential gradient. In contrast to bosonic counterpart, fermionic  rectifier allows us to control the heat current and switching effects in a much more efficient way, even in presence of tiny temperature difference. Most remarkably, we identify  universal nature of complete, partial and no rectification conditions that are valid for all Coulomb coupled fermionic diodes.

	\par The present work is organized as follows: We introduce model and dynamics in Sec.\ref{Sec. II}, and the steady state heat current in Sec.~\ref{Sec-III}. Microscopic picture behind the thermodynamically consistent heat flow direction is summarized in Sec.~\ref{Sec-IV}. While universal characteristics based on two dimensionless parameters are presented in Sec.~\ref{magic mean}, the role of efficient thermal switch and ideal  rectification effects are discussed in Sec.~\ref{HCM}. Finally, we conclude in Sec.~\ref{Con}.  
	
	\section{Model and Dynamics}\label{Sec. II}
	
	\begin{figure}
		\centering
		\includegraphics[width=0.85\columnwidth,height=2.25cm]{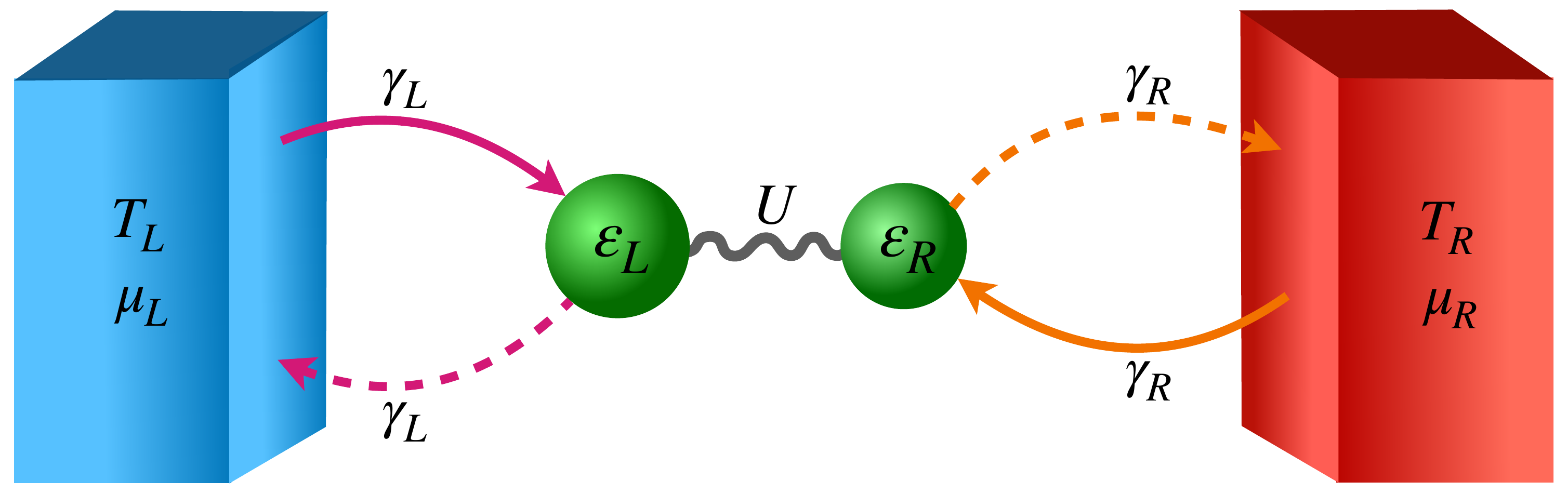}
		\caption{Coulomb coupled QDs connected with fermionic reservoirs through sequential tunnelling.}
		\label{fig:1}
	\end{figure}  
	
	\par Our model consists of two quantum dots (QDs) which are strongly and capacitively coupled to each other and interact through a long-range Coulomb force so that they can exchange energy but no particles. We consider this model following Ref~\cite{APL2017,PRBDIODE2011,zhang2018EPL,NJP2022,PRB2020} which have been recently introduced to study the thermal diode and transistor effects in a wide variety of Coulomb blockade quantum dot devices. We further assume that each QD is tunnel-coupled to its fermionic reservoir~\cite{ghosh2012PRE}. While electron transport between QDs are forbidden due to Coulomb blockade~\cite{APL2017},  electron tunnelling between QDs and its respective reservoir permits heat flow from one reservoir to another through the coupled QD system. Within sequential tunnelling~\cite{zhang2018EPL} under the Coulomb blockade regime,  each QD can have only two-levels with occupation number either zero or one.  The two QDs, as well as the temperature and chemical potential of the baths to which they are connected, are labelled by indices $L$ and $R$ [Fig.~\ref{fig:1}]. The Hamiltonian of the coupled QDs is then given by,
	\begin{eqnarray}\label{Hamiltonian}
	H_D=\sum_{\alpha=L,R} \varepsilon_\alpha d^\dagger_\alpha d_\alpha + \sum_{\alpha,\beta=L,R;\alpha\neq\beta} U_{\alpha\beta} d^\dagger_\alpha d_\alpha  d^\dagger_\beta d_\beta. 
	\end{eqnarray}
	Here, $\varepsilon_{\alpha}$ is the lowest single-particle energy of QDs. Without loss of any generality, we further assume $\varepsilon_L<\varepsilon_R$. Here $U_{\alpha\beta}$ is the positive Coulomb interaction energy between the electrons in different QDs, and $d^\dagger_\alpha(d_\alpha)$ denotes the creation (annihilation) operator for the $\alpha$-th QD,  whose eigenstates are $|0\rangle$ and $|1\rangle$ with eigenvalues $0$ and $\varepsilon_\alpha$ respectively. Since the interaction energy in Eq.~\eqref{Hamiltonian} is diagonal in the eigenbasis of the individual QDs, eigenstates of $H_D$ w.l.o.g can be written in terms of the eigenstates of the two QDs in decreasing energy order,  as $|1\rangle=|00\rangle,|2\rangle=|10\rangle,|3\rangle=|01\rangle,|4\rangle=|11\rangle$. For details we refer to Ref~\cite{APL2017,PRBDIODE2011,zhang2018EPL}. The Hamiltonian of the $\alpha$-th fermionic reservoir is defined as $H_{R}^{\alpha}=\sum_k (\varepsilon_k-\mu_\alpha)c_{\alpha k}^\dagger c_{\alpha k}$ ~\cite{NJP2022}, where $\varepsilon_k$ is the energy of the non-interacting reservoir electrons with continuous wave number $k$, $\mu_{\alpha}$ being the chemical potential and $c^\dagger(c)$ represents the creation (annihilation) operator of the electron reservoir. The coupling between QDs and the respective reservoir is described by the tunnelling Hamiltonian, $H_T^\alpha=\sum_k (t_{\alpha k} c_{\alpha k}^\dagger d_\alpha+t_{\alpha k}^* d_\alpha^\dagger c_{\alpha k})$, where $t_{\alpha k}$ is the tunnelling amplitude. Interaction within sequential tunnelling approximation imposes restriction on simultaneous tunnelling of more than one electrons at a time \cite{APL2017,NJP2022,PRBDIODE2011,PRB2020}. Consequently, there are in total four authorized transitions: the left reservoir ($L$) induces transitions between $1 \leftrightarrow 2$ and $3 \leftrightarrow 4$, while right bath ($R$) drives transitions between $1 \leftrightarrow 3$ and $2 \leftrightarrow 4$. We define transition energies $\omega_{ij}=\epsilon_i -\epsilon_j$, for $i>j$, where $\epsilon_k$ is the eigenvalue of $H_D$ for the eigenstate $|k\rangle$. In the present case, they read as $\omega_{21}=\varepsilon_L$, $\omega_{42}=\varepsilon_R+U$, $\omega_{43}=\varepsilon_L+U$, and $\omega_{31}=\varepsilon_R$. The rates at which the above transitions occur are computed using Lindblad master equation~\cite{Book1bre2002,carmichael2002book}.

	\par We implement the strong-coupling formalism following Ref.~\cite{PREREC2014,PRLQTT2016,katz2016entropy,
		jiang2015PRB,goury2019APL,liu2022entropy,gupt2022PRE} to arrive at the master equation describing the time-evolution of the density matrix $\rho$ of the coupled QDs system (See Appendix~\ref{Appendix-A})
	\begin{equation}\label{Master Eqn}
	\frac{d\rho}{dt}=-\frac{i}{\hbar}[H_D,\rho]+\mathcal{L}_L[\rho]+\mathcal{L}_R[\rho],
	\end{equation}
	under Born, Markov and secular approximation \cite{PREREC2014,PRB2018REC,Book1bre2002}. It is important to note here that the strong coupling formalism refers to the coupling between the dots, while the the system-bath coupling is still assumed to be weak, so that Born-Markov approximation can safely be implemented \cite{PREREC2014,PRLQTT2016,katz2016entropy,jiang2015PRB,goury2019APL,liu2022entropy,gupt2022PRE}. This implies that the Lindbladians are obtained on the basis of the eigenstates of the full system Hamiltonian $H_D$. Thus dissipation mechanism of each QD depends not only on the coupling to its own bath, but also on the coupling between QDs, which is necessary for accurately describing the heat flow and rectification effects over a wide range of system parameters as considered below.
	
	\section{Evaluation of Steady State Heat current}\label{Sec-III}
	\par
\begin{figure*}[t]
	\centering
	\includegraphics[width=15cm,height=8cm]{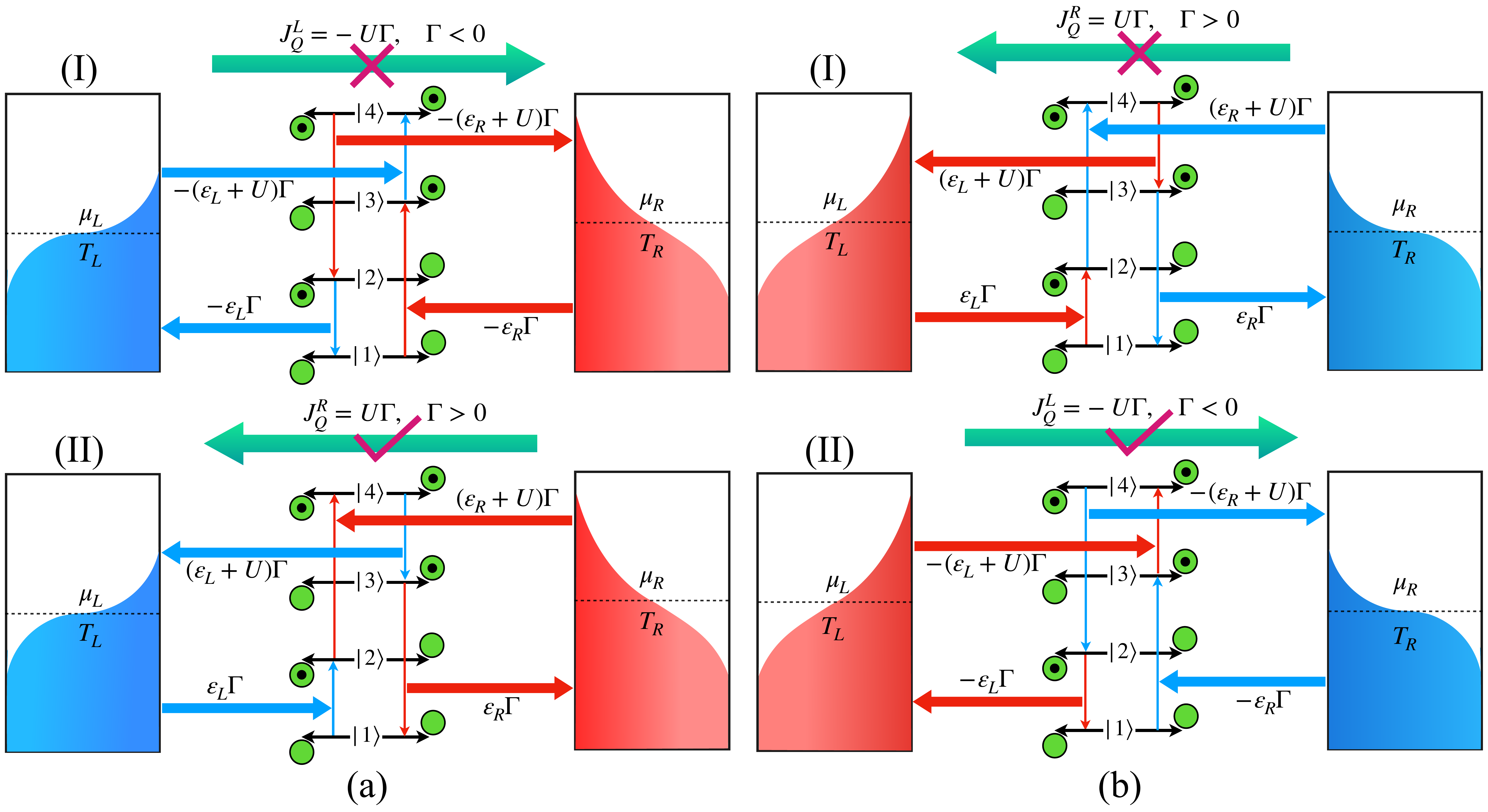}
	\caption{Thermodynamically consistent and inconsistent heat flow directions: Probable paths for transition cycle under (a) $T_L<T_R$ ; (b) $T_R<T_L$. In both conditions, Path-I is initiated by the hot bath and Path-II is initiated by the cold bath. Though Path-I seems to be natural heat flow direction, transition cycle actually encompasses Path-II in both limits, following the laws of thermodynamics.}
	\label{fig:App1}
    \end{figure*}

 In the present model particles can not be exchanged between QDs as they interact only through long range Coulomb force. As a result, there is no particle flow in between reservoirs through the QD system~\cite{APL2017,PRBDIODE2011,zhang2018EPL,NJP2022,PRB2020}. So, energy is exchanged only in the form of heat [See Figs. ~\ref{fig:App1} and ~\ref{fig:App2}]. The expression of the heat current can then be obtained using the standard procedure \cite{david2015Elsevier} starting from the von-Neumann entropy of the system, defined as $\mathcal{S}[\rho(t)]=-k_B{\rm Tr}[\rho(t)\ln\rho(t)]$. Upon taking the time derivative of the von-Neumann entropy, one obtains
	\begin{equation}\label{dot-von-neumann-entropy}
	\frac{d}{dt}\mathcal{S}[\rho(t)]=-\sum_{\alpha=L,R} k_B {\rm Tr}\{\mathcal{L}_{\alpha}[\rho(t)]\ln\rho(t)\},
	\end{equation}
	where we have used the master equation~\eqref{Master Eqn} and ${\rm Tr}[\dot{\rho}]=0$. Sphon inequality~\cite{spohn1978JMP} in the form of second law of thermodynamics~\cite{kosloff2013Entropy} for any Lindblad super-operator $\mathcal{L}$ can be written as ${\rm Tr}\{\mathcal{L}[\rho(t)](\ln\rho(t)-\ln\rho_{ss})\}\leqslant0$. Here $\rho_{ss}$ is the steady state population of the system, satisfying $\mathcal{L}[\rho_{ss}]=0$. Denoting the stationary state of $\mathcal{L}_{\alpha}$, as $\rho_{ss}^\alpha$, above inequality can be applied to both the terms of the sum in Eq.~\eqref{dot-von-neumann-entropy}, yielding
	\begin{equation}\label{spohn-inequality-dot}
	\sum_{\alpha=L,R} {\rm Tr}\{\mathcal{L}_{\alpha}[\rho(t)](\ln\rho(t)-\ln\rho_{ss}^\alpha)\}\leqslant0.
	\end{equation}
	From Eqs.~\eqref{dot-von-neumann-entropy} and~\eqref{spohn-inequality-dot} we can write
	\begin{equation}\begin{split}\label{spohn-inequality}
	\frac{d}{dt}\mathcal{S}[\rho(t)]+\sum_{\alpha=L,R}k_B{\rm Tr}[\mathcal{L}_{\alpha}[\rho(t)]\ln\rho_{ss}^\alpha]\geq0
	\end{split}.
	\end{equation}
	Above equation can be compared with the dynamical version of the second law given by \cite{kosloff2013Entropy,david2015Elsevier}
	\begin{equation}\label{dynamical-2nd-law}
	\frac{d}{dt}\mathcal{S}[\rho(t)]-\sum_{\alpha=L,R} \frac{J_Q^\alpha(t)}{T_\alpha} \geq0.
	\end{equation}
	This allows us to identify the heat current (energy flow rate) $J_Q^\alpha(t)$ associated with the $\alpha$-th reservoir as
	\begin{equation}\label{heat-current-gen}
	J_Q^\alpha(t)=-\frac{1}{\beta_\alpha} {\rm Tr}\{\mathcal{L}_\alpha[\rho(t)]\ln{\rho_{ss}^\alpha}\}.
	\end{equation}
	
	For the steady state operation, the master equation~\eqref{Master Eqn} drives the system towards a Gibbs-like stationary state, characterized by \cite{david2015Elsevier,gupt2022PRE}
	\begin{equation}\label{rho_ss}
	\rho_{ss}^\alpha=\mathcal{Z}^{-1}\exp[-\beta_\alpha H_D],
	\end{equation}
	where $\mathcal{Z}= {\rm Tr}\{\exp[{-\beta_\alpha H_D}]\}$. For the detailed derivation of the steady state heat current we refer to the review articles by Kosloff and Gelbwaser et al. \cite{kosloff2013Entropy,david2015Elsevier}. Inserting Eq.~\eqref{rho_ss} into Eq.~\eqref{heat-current-gen}, we obtain the general expression for the heat current, which under stead state condition simplifies to  
	\begin{equation}\label{heat-current-final}
	J_Q^\alpha={\rm Tr}\{\mathcal{L}_\alpha[\rho_{ss}]H_D\}=\sum_{\omega_{ij}}\omega_{ij}\Gamma^{\alpha}_{ij}.
	\end{equation}
	Here the net decaying rate $\Gamma_{ij}^{\alpha}$ from $|i\rangle$ to $|j\rangle$ ($i>j$) is denoted as
	\begin{equation}\begin{split}\label{gamma_ij}
	\Gamma_{ij}^{\alpha}=&-\Gamma_{ji}^{\alpha}
	\\\equiv&\gamma_{\alpha}[1-f_{\alpha}(\omega_{ij})]\rho_{ii}-\gamma_{\alpha}f_{\alpha}(\omega_{ij})\rho_{jj}=[\Gamma_{ij}^{\alpha}]_\downarrow-[\Gamma_{ij}^{\alpha}]_\uparrow.
	\end{split}
	\end{equation}
	The first term represents the emission and the second term corresponds to absorption. $\gamma_\alpha$ being the bare tunnelling rate between dots and respective reservoirs [Fig.~\ref{fig:1}]. $f_\alpha(\omega_{ij})$ is the Fermi Distribution function (FDF) 
	\begin{equation}\begin{split}
	f_\alpha(\omega_{ij})\equiv f_\alpha(\omega_{ij},T_\alpha,\mu_\alpha)=\bigg[{1+\exp \bigg({\frac{\omega_{ij}-\mu_\alpha}{k_BT_\alpha}}}\bigg)\bigg]^{-1} 
	\end{split},
	\end{equation}
	corresponding to the transition energy $\omega_{ij}=\epsilon_i-\epsilon_j$ between eigenstates $\rvert i\rangle$ and $\rvert j\rangle$ controlled by the $\alpha$-th reservoir. Tunnelling of electrons into or out of QDs are primarily governed by FDF. In our model there are four allowed transitions and both lead guides two transitions each: $L$ lead drives transitions between $\rvert 1\rangle\leftrightarrow\rvert 2\rangle$ and $\rvert 4\rangle\leftrightarrow\rvert 3\rangle$, while $R$ lead controls $\rvert 1\rangle\leftrightarrow\rvert 3\rangle$ and $\rvert 4\rangle\leftrightarrow\rvert 2\rangle$ transitions. For the sake of convenience of our analysis, corresponding FDFs are expressed in terms of two dimensionless parameters, effective tunnelling barrier ($\chi_{\alpha}={(\varepsilon_{\alpha}-\mu_{\alpha})}/{U}$) and dimensionless thermal energy ($\xi_{\alpha}={k_B T_{\alpha}}/{U}$) as follows
	\begin{widetext}
	\begin{equation}\begin{split}\label{FDF-expressions}
	f_L(\omega_{21})=\left[{1+\exp \bigg({\frac{\varepsilon_L-\mu_L}{k_BT_L}}\bigg)}\right]^{-1}=\left[{1+\exp \bigg({\frac{\chi_L}{\xi_L}}\bigg)}\right]^{-1}=&f_L^1,
	\\f_L(\omega_{43})=\left[{1+\exp \bigg({\frac{\varepsilon_L+U-\mu_L}{k_BT_L}}\bigg)}\right]^{-1}=\left[{1+\exp \bigg({\frac{\chi_L+1}{\xi_L}}\bigg)}\right]^{-1}=&f_L^2,
	\\f_R(\omega_{31})=\left[{1+\exp \bigg({\frac{\varepsilon_R-\mu_R}{k_BT_R}}\bigg)}\right]^{-1}=\left[{1+\exp \bigg({\frac{\chi_R}{\xi_R}}\bigg)}\right]^{-1}=&f_R^1,
	\\f_R(\omega_{42})=\left[{1+\exp \bigg({\frac{\varepsilon_R+U-\mu_R}{k_BT_R}}\bigg)}\right]^{-1}=\left[{1+\exp \bigg({\frac{\chi_R+1}{\xi_R}}\bigg)}\right]^{-1}=&f_R^2.
	\end{split}
	\end{equation}
	\end{widetext}

	\par Now, our task is to first evaluate the full expression of $\Gamma_{ij}^{\alpha}$ at the steady state and then, find out the expression of the heat current. Under the steady state condition master Eq.~\eqref{Master Eqn} is characterized by $\dot{\rho}_{ss}=0$ which reduces to 
	\begin{equation}\begin{split}\label{SS}
	&\dot{\rho}_{11}=0=\Gamma_{31}^R -\Gamma_{12}^L; \quad \quad \dot{\rho}_{22}=0=\Gamma_{12}^L -\Gamma_{24}^R,
	\\&\dot{\rho}_{33}=0=\Gamma_{43}^L -\Gamma_{31}^R; \quad \quad \dot{\rho}_{44}=0=\Gamma_{24}^R -\Gamma_{43}^L. \ \  \end{split}
	\end{equation}
	Thus, at the steady state, all the net transitions rates become equal to $\Gamma$, i.e., $\Gamma_{31}^R=\Gamma_{12}^L=\Gamma_{24}^R=\Gamma_{43}^L\equiv\Gamma$. The four set of equations in Eq.~\eqref{SS} are not independent since $\sum_i \rho_{ii}=1$, which uniquely solves all the state occupation probabilities as well as the heat current in terms of a single quantity $\Gamma$:\ $\Gamma_{31}^R=\Gamma_{12}^L=\Gamma_{24}^R=\Gamma_{43}^L\equiv\Gamma$. From Eq.~\eqref{heat-current-final}, we can then evaluate the general expression of heat current at the steady state as follows
	\begin{equation}\begin{split}\label{J-explicit}
	J_Q^R=\varepsilon_R\Gamma_{13}^R-(\varepsilon_R+U)\Gamma_{42}^R=&-\varepsilon_R\Gamma+(\varepsilon_R+U)\Gamma=U\Gamma,
	\\ J_Q^L=\varepsilon_L\Gamma_{12}^L-(\varepsilon_L+U)\Gamma_{43}^L=& \ \varepsilon_L\Gamma-(\varepsilon_L+U)\Gamma=-U\Gamma.
	\end{split}
	\end{equation}
	Thus we finally arrive at the explicit analytical expression of the steady state heat current as
	\begin{equation}\begin{split}\label{J}
	J_Q^L=-J_Q^R=-U\Gamma.
	\end{split}
	\end{equation}
	To find out $\Gamma$, we rewrite Eq.~\eqref{SS} in terms of $f_{L(R)}^{1}$ and $f_{L(R)}^{2}$ 
	\begin{widetext}
	\begin{equation}\begin{split}\label{C7}
	\dot{\rho}_{11}=\gamma_R[1-f_R^1]\rho_{33}-[\gamma_Rf_R^1+\gamma_Lf_L^1]\rho_{11}+\gamma_L[1-f_L^1]\rho_{22}=0,&
	\\\dot{\rho}_{22}=\gamma_Lf_L^1\rho_{11}-[\gamma_L(1-f_L^1)+\gamma_Rf_R^2]\rho_{22}+\gamma_R[1-f_R^2]\rho_{44}=0,&
	\\\dot{\rho}_{33}=\gamma_Rf_R^1\rho_{11}-[\gamma_Lf_L^2+\gamma_R(1-f_R^1)]\rho_{33}+\gamma_L[1-f_L^2]\rho_{44}=0,&
	\\\dot{\rho}_{44}=\gamma_Lf_L^2\rho_{33}-[\gamma_R(1-f_R^2)+\gamma_L(1-f_L^2)]\rho_{44}+\gamma_Rf_R^2\rho_{22}=0.&
	\end{split}
	\end{equation}
	defined through Eq.~\eqref{FDF-expressions} and find out the steady state populations subject to the condition
	\begin{equation}\label{C8}
	\rho_{11}+\rho_{22}+\rho_{33}+\rho_{44}=1.
	\end{equation} 
	Using Eq.~\eqref{C7} and Eq.~\eqref{C8} we can construct 
	\begin{equation}\label{C9}
	\mathcal{M}\begin{bmatrix}
	\rho_{11}
	\\\rho_{22}
	\\\rho_{33}
	\\\rho_{44}
	\end{bmatrix}=\begin{bmatrix}
	0
	\\0
	\\0
	\\1
	\end{bmatrix},
	\end{equation}
	where,
	%\begin{widetext}
		\begin{equation}
	\mathcal{M}=\begin{bmatrix}
	-[\gamma_Rf_R^1+\gamma_Lf_L^1] & \gamma_L[1-f_L^1] & \gamma_R[1-f_R^1] & 0
	\\\gamma_Lf_L^1 & [\gamma_L(1-f_L^1)+\gamma_Rf_R^2] & 0  & \gamma_R[1-f_R^2]
	\\\gamma_Rf_R^1 & 0 & [\gamma_Lf_L^2+\gamma_R(1-f_R^1)] & \gamma_L[1-f_L^2]
	\\ 1 & 1 & 1 & 1
	\end{bmatrix}.
	\end{equation}
	Solving above equation, expressions for the steady state populations $\rho_{ii}$ are found to be
	\begin{equation*}\begin{split}
	\rho_{11}=-\frac{1}{|\mathcal{M}|} [&\gamma_L{\gamma_R}^2\{(1-f_L^1)(1-f_R^1)(1-f_R^2)+(1-f_L^2)(1-f_R^1)f_R^2\}\\
	+&\gamma_R{\gamma_L}^2\{(1-f_L^1)(1-f_R^2)f_L^2+(1-f_L^2)(1-f_R^1)(1-f_L^1)\}],\\
	\rho_{22}=-\frac{1}{|\mathcal{M}|} [&\gamma_L{\gamma_R}^2\{f_L^1(1-f_R^1)(1-f_R^2)+f_L^2(1-f_R^2)f_R^1\}\\
	+&\gamma_R{\gamma_L}^2\{f_L^1(1-f_R^2)f_L^2+(1-f_L^2)(1-f_R^1)f_L^1\}],\\
	\rho_{33}=-\frac{1}{|\mathcal{M}|} [&\gamma_L{\gamma_R}^2\{(1-f_L^2)f_R^1f_R^2+(1-f_L^1)(1-f_R^2)f_R^1\}\\
	+&\gamma_R{\gamma_L}^2\{f_R^2(1-f_L^2)f_L^1+(1-f_L^2)f_R^1(1-f_L^1)\}],\\
	\rho_{44}=-\frac{1}{|\mathcal{M}|} [&\gamma_L{\gamma_R}^2\{f_L^1f_R^2(1-f_R^1)+f_L^2f_R^1f_R^2\}\\
	+&\gamma_R{\gamma_L}^2\{f_L^1f_R^2f_L^2+(1-f_L^1)f_L^2f_R^1\}],
	\end{split}
	\end{equation*}
	where $|\mathcal{M}|$ stands for the determinant of the matrix $\mathcal{M}$.
	Using the above expressions we can evaluate the final expression of $\Gamma$ as follows
		\begin{equation}\label{C10}
	\Gamma=\frac{\gamma_L+\gamma_R}{\gamma_L\gamma_R}\left[\frac{f_R^1f_R^2f_L^2-f_R^1f_R^2f_L^1+f_L^1f_L^2f_R^1-f_L^1f_L^2f_R^2+f_L^1f_R^2-f_L^2f_R^1}{f_L^1f_R^1+f_L^2f_R^2-f_L^1f_R^2-f_L^2f_R^1-1}\right].
	\end{equation}
	Using Eq.~\eqref{C10}, final expression of the steady state heat current in terms of dimensionless parameters $\chi_{\alpha}$ and $\xi_{\alpha}$ can be written as
	\begin{equation}\label{J2}
	{J}^L_Q=-{J}^R_Q=-\frac{U(\gamma_L+\gamma_R)}{\gamma_L\gamma_R}\left[\frac{\exp\bigg({\frac{1}{\xi_L}}\bigg)-\exp\bigg({\frac{1}{\xi_R}}\bigg)}{X}\right],%\nonumber,
	\end{equation}
	where,
	\begin{equation}\begin{split}\label{J-final}
	X=&\bigg[\exp\bigg({\frac{1}{\xi_L}\bigg)}\left[2+\exp\bigg({-\frac{\chi_R}{\xi_R}\bigg)}+\exp\bigg({\frac{\chi_L}{\xi_L}\bigg)}+\exp\bigg({\frac{\chi_L}{\xi_L}-\frac{\chi_R}{\xi_R}}\bigg)\right]
	\\+&\exp\bigg({\frac{1}{\xi_R}}\bigg)\left[2+\exp\bigg({-\frac{\chi_L}{\xi_L}}\bigg)+\exp\bigg({\frac{\chi_R}{\xi_R}}\bigg)+\exp\bigg({\frac{\chi_R}{\xi_R}-\frac{\chi_L}{\xi_L}}\bigg)\right]
	\\+&\exp\bigg({\frac{1}{\xi_L}+\frac{1}{\xi_R}}\bigg)\left[\exp\bigg({\frac{\chi_R}{\xi_R}}+\exp{\frac{\chi_L}{\xi_L}}\bigg)+\exp\bigg({\frac{\chi_L}{\xi_L}+\frac{\chi_R}{\xi_R}}\bigg)\right]
	\\+&\left[\exp\bigg({-\frac{\chi_R}{\xi_R}}\bigg)+\exp\bigg({-\frac{\chi_L}{\xi_L}}\bigg)+\exp\bigg({-\frac{\chi_L}{\xi_L}-\frac{\chi_R}{\xi_R}}\bigg)\right]\bigg]. 
	\end{split}
	\end{equation}
    \end{widetext}
	
	This is the exact analytical expression of heat current derived under Born-Markov master equation. Since $U$, $\gamma_L$, $\gamma_R$ and $X$ are all positive, it is immediately clear from Eqs.~\eqref{J2}-\eqref{J-final}, that if $\xi_R>\xi_L$ ($T_R>T_L$), heat will flow from right to left ($J_R>0$) in accordance with Eq.~\eqref{J}.

	\section{Microscopic description of heat flow}\label{Sec-IV}

	\par Classically heat flows according to laws of thermodynamics from high to low temperature. Quantum mechanically, flow of heat current must be governed by a microscopic description, without violating the ultimate laws of thermodynamics~\cite{kosloff2013Entropy,deffner2019Book,Ghosh2019,gupt2021PRE,ghosh2013PRE}. 
	Since the QDs are strongly coupled with each other, they form a four level system as depicted in Fig.~\ref{fig:App1}, where inter-dot transition  is restricted due to long range Coulomb force. So, the energy transfer is only caused by the heat baths via the coupled states of the overall system~\cite{APL2017,PRBDIODE2011,zhang2018EPL,NJP2022,PRB2020}.For $T_R>T_L$, natural heat flow direction will be from $R$ to $L$ with $J_Q^R >0$. In this case, the first excitation from the ground state must be guided by the cold bath, instead of hot bath, which may appear paradoxical at first sight but makes the net transition rate $\Gamma>0$ as required by Eq.~\eqref{J}. If the cycle follows the opposite path $1 \rightarrow 3 \rightarrow 4 \rightarrow 2 \rightarrow 1$ [Fig.~\ref{fig:App1}(a)-I], then $J_Q^L>0$, as $U\Gamma$ amount of energy should be delivered by the left bath to the right bath, which is in complete disagreement with the laws of thermodynamics. So, in order to be consistent with the laws of the thermodynamics, transition cycle must run in $1 \rightarrow 2 \rightarrow 4 \rightarrow 3 \rightarrow 1$, initiated by the cold bath. This is seemingly paradoxical in the sense that classically we expect during heat flow, energy is supplied by the hot bath and dumped into the cold bath. For the specific example, it is however more favourable for the cold bath to make the $1 \rightarrow 2$ transition in Fig.~\ref{fig:App1}(a)-II which costs $\varepsilon_L$ amount of energy than $3 \rightarrow 4$ transition in Fig.\ref{fig:App1}(a)-I which requires $(\varepsilon_L+U)$ amount of energy.  It is interesting to note that for $T_L>T_R$, (i.e., $J_Q^L>0$, heat flows from left to right following Fig.~\ref{fig:App1}(b)-II), first transition is still mediated by the cold bath between $1 \rightarrow 3$, as it requires less amount of energy ($\varepsilon_R$) than $2 \rightarrow 4$ transition ($\varepsilon_R+U$) in Fig.~\ref{fig:App1}(b)-I.
	\begin{figure}[h]
		\centering
		\includegraphics[width=\columnwidth,height=5cm]{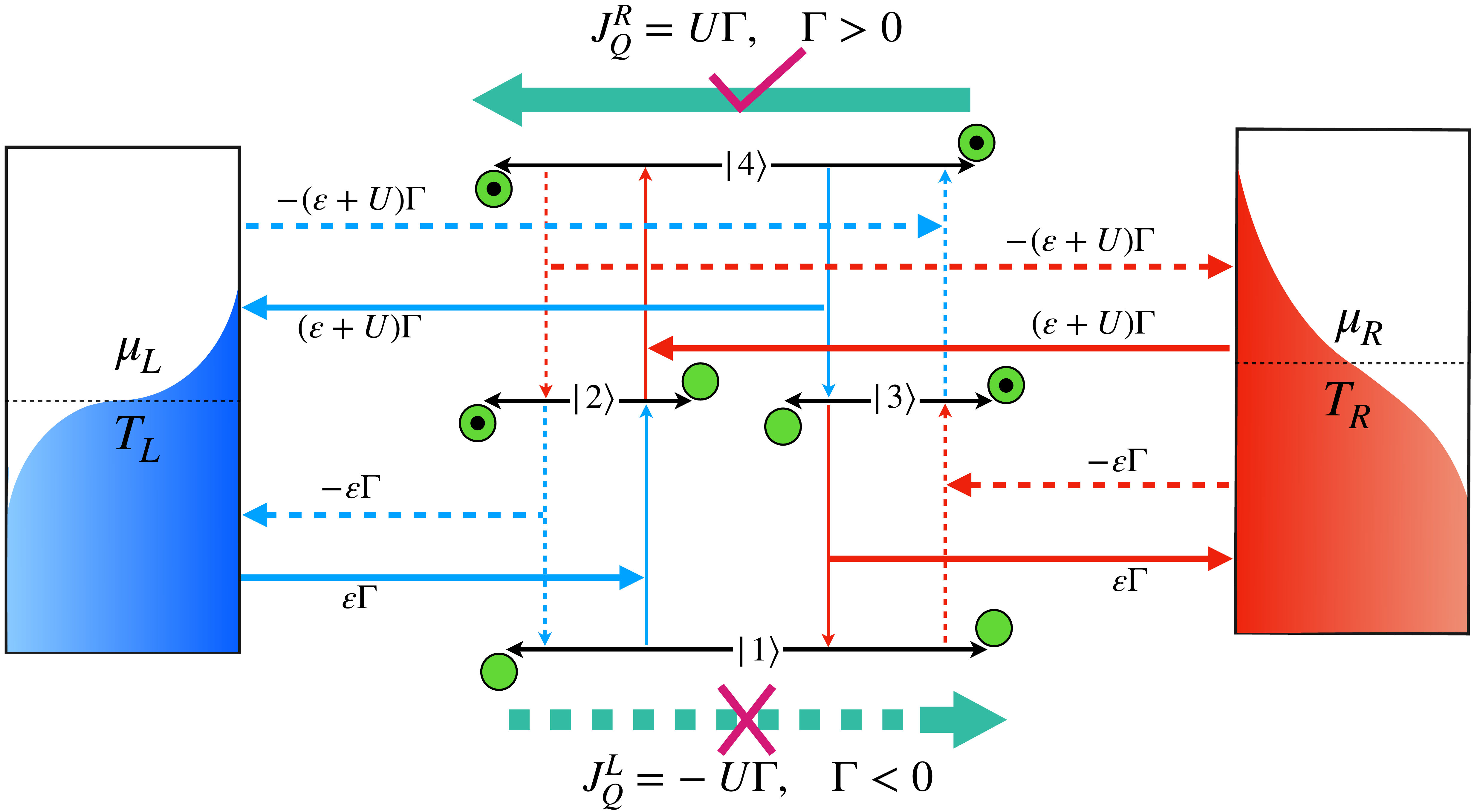}
		\caption{Thermodynamically consistent transition cycle for $T_R>T_L$ in case of $\varepsilon_L=\varepsilon_R=\varepsilon$; first excitation from $|1\rangle$ must be guided by cold bath (Path-II: solid lines), instead of hot bath (Path-I: dotted lines).}
		\label{fig:App2}
	\end{figure}
	
	It is important to emphasize that although we have taken $\varepsilon_L<\varepsilon_R$ to draw the schematic energy level diagram of Fig.~\ref{fig:App1}, magnitude of the heat current in particular depends only on the two dimensionless parameters $\{ \chi_{\alpha},\xi_{\alpha}\}$ via Eq.~\eqref{J2}. In the next section, we will explore the universal characteristics of the heat current solely based on these two dimensionless system parameters. For instance, Fig.~\ref{fig:App2} shows that basic principle behind the thermodynamically consistent transition cycle remains same even in case of $\varepsilon_L=\varepsilon_R$.

    \section{Universal characteristic due to magic mean}\label{magic mean}
    
    \par
\begin{figure}[h]
    	\centering
    	\includegraphics[width=\columnwidth,height=8.5cm]{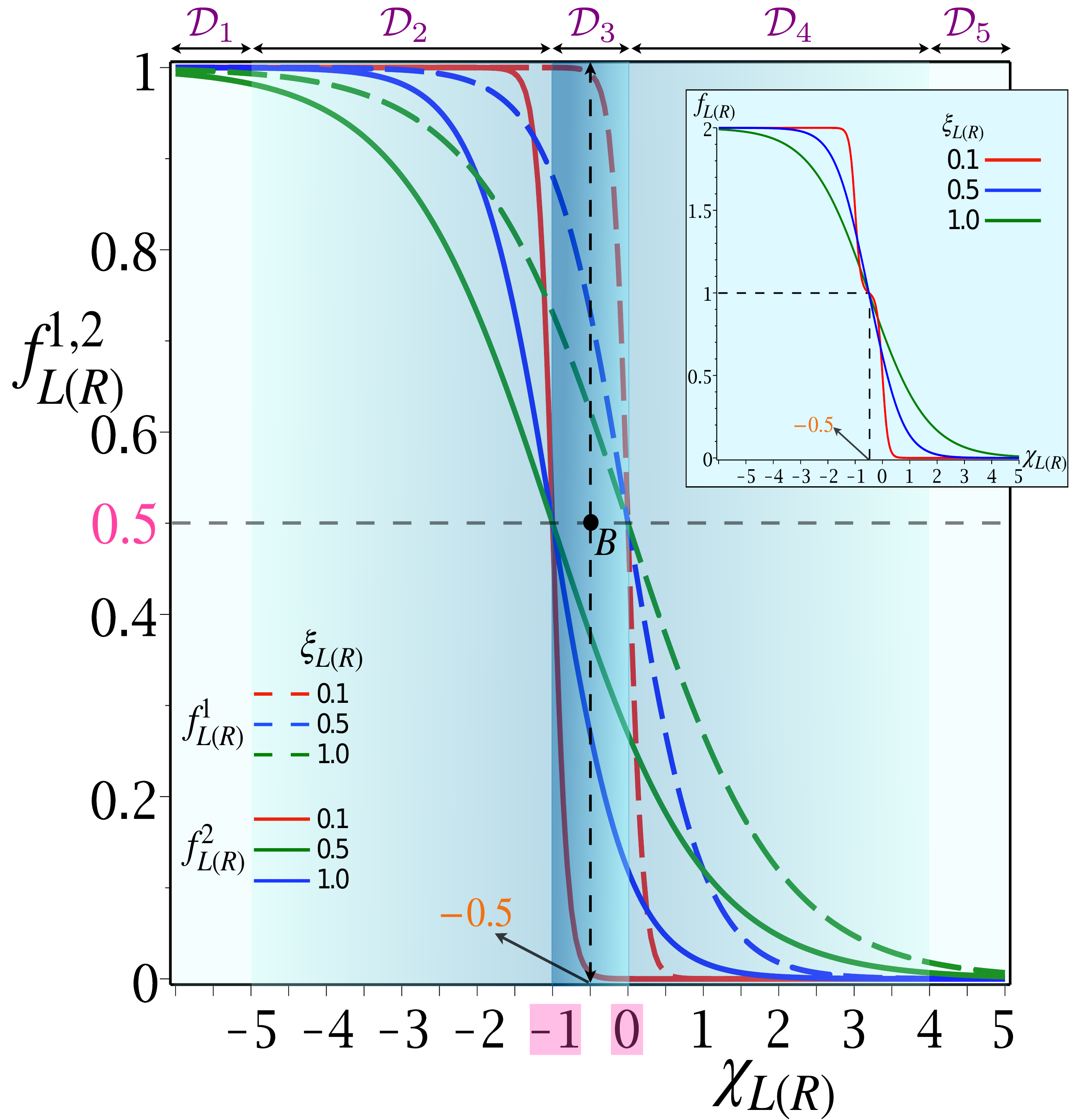}
    	\caption{$\mathcal{D}_3$ always spreads between transition points of $f_{L(R)}^{1}$ and $f_{L(R)}^{2}$; span of $\mathcal{D}_{1[5]}$ and $\mathcal{D}_{2[4]}$ are not fixed and strongly depends on the value of $\xi_{L(R)}$. Color gradients signifies the magnitude of $|J_Q|$.} 
    	\label{fig:3}
    \end{figure}

 While temperature gradient dictates the overall heat flow direction, chemical potential plays a very important role in determining the heat current magnitudes. To illustrate, we note that FDFs $f_{L(R)}^{1}$ and $f_{L(R)}^{2}$ in Eq.~\eqref{FDF-expressions}, expressed in terms of dimensionless thermal energy $\xi_{L(R)}=k_B T_{L(R)}/U$ and effective tunnelling barrier $\chi_{L(R)}=(\varepsilon_{L(R)}-\mu_{\alpha})/U$, are constrained by $0\leq f_{L(R)}^{2}<f_{L(R)}^{1} \leq 1$ and become $0.5$ at $\chi_{L(R)}=0$ and $-1$, which correspond to $\mu_{L(R)}=\omega_{21(31)}=\varepsilon_{L(R)}$ and $\mu_{L(R)}=\omega_{43(42)}=\varepsilon_{L(R)}+U$, respectively. According to Eq.~\eqref{gamma_ij}, if $0 \leq \{f_{L(R)}^{1},f_{L(R)}^{2}\} \leq 0.5$, it favours de-excitation and equivalently excitation is favoured when $0.5 \leq \{f_{L(R)}^{1},f_{L(R)}^{2}\} \leq 1$, for the corresponding transition. This is unique to fermionic reservoirs and allows us to implement a systematic analytical scheme solely based on FDFs which is in sharp contrast to its bosonic counterpart. The span of $f_{L(R)}^{1}$ and $f_{L(R)}^{2}$ as a function $\chi_{L(R)}$, can thus be classified into five domains $\mathcal{D}_{[1-5]}$ [Fig.~\ref{fig:3}: Main] and the total FDF, defined as $f_{L(R)}=f_{L(R)}^{1}+f_{L(R)}^{2}$, is found to be spreading between $0\leq f_{L(R)} \leq 2$ [Fig.~\ref{fig:3}: Inset].
    
    \par
\begin{figure*}[t]
    	\centering
    	\includegraphics[width=0.85\columnwidth,height=7.5cm]{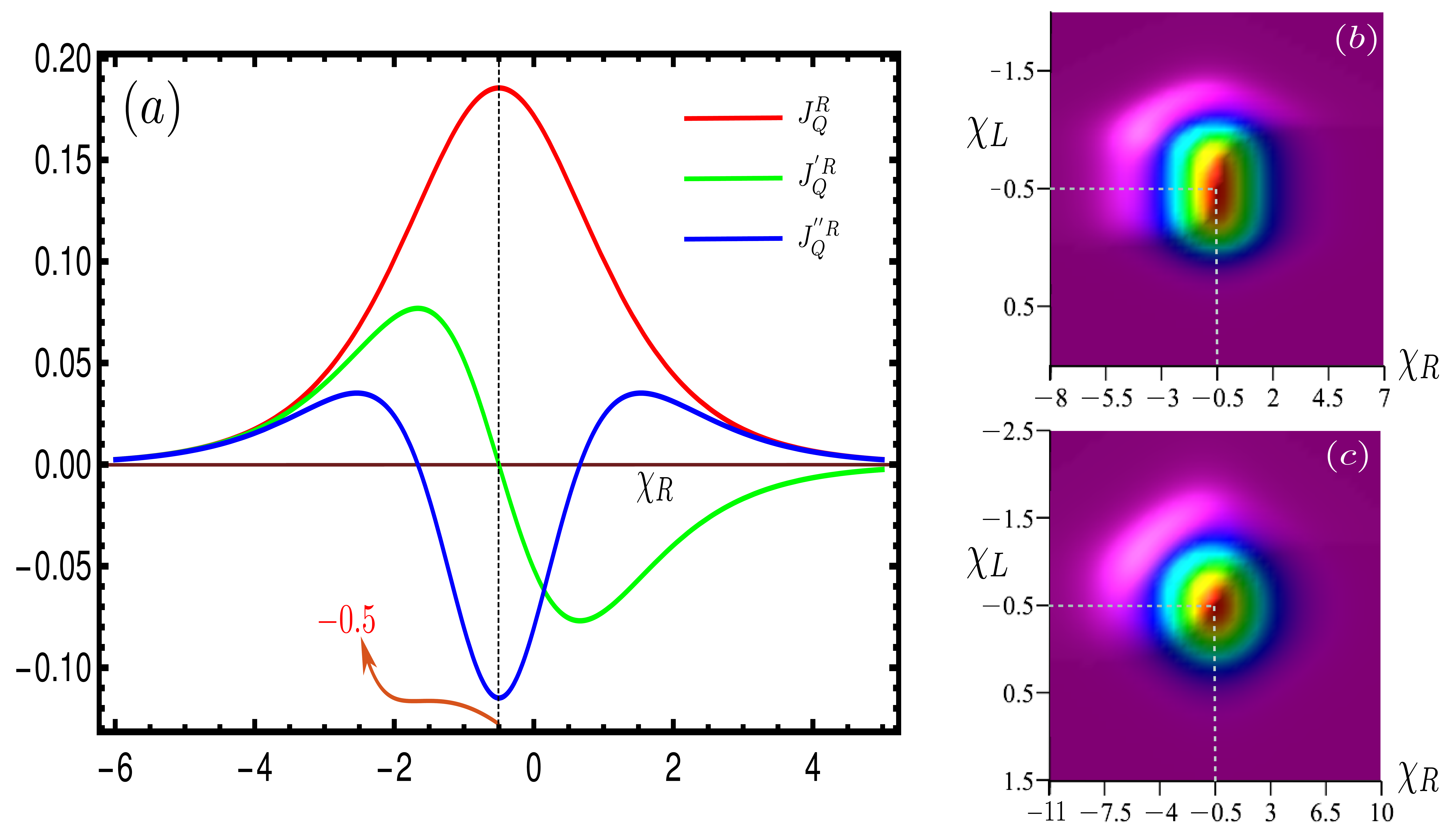}
    	\caption{(a) Variation of heat current $J_Q^R$ (red line) with $\chi_R$ for fixed $\chi_L$, $\xi_R$ and $\xi_L$. Heat current become maximum at $\chi_R=-0.5$ which is supported by the plots of first and second derivatives of heat current: $J_Q^{\prime R}$ (green line) is zero and $J_Q^{\prime \prime R}$ (blue line) is negative at the magic mean point $\chi_R=-0.5$. Absolute value of the heat current $\arrowvert J_Q\rvert$ is plotted as a function of (b) $\{\chi_{L},\chi_{R}\}$ for $\xi_R=0.5$ and $\xi_L=0.2$; (c) $\{\chi_{L},\chi_{R}\}$ for $\xi_R=1$ and $\xi_L=0.5$ }
    	\label{fig derivative}
    \end{figure*}

 In domain $\mathcal{D}_{1[5]}$, both $f_{L(R)}^{1}$ and $f_{L(R)}^{2}\sim1[0]$, thus Eq.~\eqref{gamma_ij} reduces to $\Gamma_{21/43}^L\simeq\{\Gamma_{21/43}^L\}_{\uparrow[\downarrow]}$ and $\Gamma_{31/42}^R\simeq\{\Gamma_{31/42}^R\}_{\uparrow[\downarrow]}$ i.e. only absorption (emission) is allowed in all four transitions, hence transition cycle can\textquotesingle t be completed [Fig.~\ref{fig:App1}] which results in vanishing $\rvert J_Q\arrowvert \rightarrow 0$. Now, $\mathcal{D}_{2[4]}$ is characterised by $0.5[0]\lesssim \{f_{L(R)}^{1},f_{L(R)}^{2}\} \lesssim 1[0.5]$, hence excitation as well as de-excitation is favoured for all four transitions yielding non-zero heat current. On the contrary, domain $\mathcal{D}_3$, parametrized by $-1\leq\chi_{L(R)}\leq0$, is sharply defined between the transition points $f_{L(R)}^2$ and $f_{L(R)}^1$. As opposed to $\mathcal{D}_{3}$, both $f_{L(R)}^{1}$ and $f_{L(R)}^{2}$ are closer to $1[0]$ in $\mathcal{D}_{2[4]}$, so that heat flux decreases in $\mathcal{D}_{2[4]}$ relative to $\mathcal{D}_3$. In $\mathcal{D}_3$, both $f_{L(R)}^1$ and $f_{L(R)}^2$ take values such that excitation from $\lvert1\rangle \rightarrow\lvert2\rangle (\lvert2\rangle \rightarrow\lvert4\rangle)$ and de-excitation from $\lvert4\rangle \rightarrow\lvert3\rangle (\lvert3\rangle \rightarrow\lvert1\rangle)$ can occur simultaneously at an optimal rate. Condition becomes ideal at the midpoint of $\mathcal{D}_3$ (point $B$ in Fig.~\ref{fig:3}) which corresponds to $\chi_{L(R)}=-0.5$. To find out analytically, the optimal value of $\chi_{L(R)}$ for which $\arrowvert {J_Q}\rvert$ becomes maximum, first we have to differentiate ${J_Q} \equiv J_Q^R$ (Cf.~Eq.~\eqref{J2}) w.r.t. $\chi_{L(R)}$ and set that equal to be zero. Now, by differentiating Eq.~\eqref{J2} w.r.t $\chi_{R}$, for fixed $\{\chi_L,\xi_L,\xi_R\}$, we obtain,
    \begin{widetext}
    	\begin{equation}\begin{split}\label{A17}
    &\left(\frac{\partial J_Q^R}{\partial\chi_R}\right)_{\chi_L,\xi_L,\xi_R}\\
    &=\frac{U(\gamma_L+\gamma_R)}{\gamma_L\gamma_R}\left[\frac{\partial}{\partial\chi_R}\left\{\frac{f_R^1f_R^2f_L^2-f_R^1f_R^2f_L^1+f_L^1f_L^2f_R^1-f_L^1f_L^2f_R^2+f_L^1f_R^2-f_L^2f_R^1}{f_L^1f_R^1+f_L^2f_R^2-f_L^1f_R^2-f_L^2f_R^1-1}\right\}\right] \\
    &= 0.
    \end{split}
    \end{equation}
    \end{widetext}
    As, $U>0$ and $\gamma_{L},\gamma_{R}\neq0$, Eq.~\eqref{A17} implies that the only non-trivial solution for maximizing $|J_Q|$ is equivalent to maximizing $\Gamma$, which is given by the criteria
    \begin{equation}\label{A19}
    f_R^1+f_R^2=1.
    \end{equation}
    It is clear from Eq.~\eqref{A17} that heat current vanishes under two limiting conditions: $(i)\ f_R^1=f_R^2=0;\; (ii)\ f_R^1=f_R^2=1$. Again from Eq.~\eqref{FDF-expressions}, we can write 
    \begin{equation}\label{A21}	
    1\geq f_\alpha^1\geq f_\alpha^2 \geq0, \quad \alpha=L,R.
    \end{equation}
    So, Eq.~\eqref{A21} signifies that, if $f_\alpha^1=0$ then $f_\alpha^2$ is certainly $0$. Similarly, if $f_\alpha^2=1$ then $f_\alpha^1$ is certainly $1$. In both cases, heat current vanishes. Thus, the only condition for nonzero heat current reduces to $f_R^1\neq0$ and $f_R^2\neq1$. Under these conditions, after solving Eq.~\eqref{A19}, we get the value of $\chi_R=-0.5$, for which $|J_Q|$ attains maximum:
    \begin{widetext}
    	\begin{equation}\begin{split}
    \ &f_R^1=1-f_R^2
    \\ \text{or},\ &\frac{1}{f_R^1}=\frac{1}{1-f_R^2}
    \\ \text{or},\ &1+\exp\bigg({\frac{\chi_R}{\xi_R}}\bigg)=1+\exp\bigg({-{\frac{\chi_R+1}{\xi_R}}}\bigg)
    \\ \text{or},\ &\exp\bigg({\frac{\chi_R}{\xi_R}}\bigg)=\exp\bigg({-{\frac{\chi_R+1}{\xi_R}}}\bigg)
    \\ \text{or},\ &\frac{\chi_R}{\xi_R}=-{\frac{\chi_R+1}{\xi_R}}
    \\ \text{or},\ &\chi_R=-\chi_R-1
    \\ \text{or},\ &\chi_R=-0.5.
    \end{split}
    \end{equation}
    \end{widetext}
    So, heat current will be maximum when $\chi_R=-0.5$ [Fig~\ref{fig derivative}(a)]; similarly, when $\chi_L$ varies, $\chi_L=-0.5$ is the criteria for having highest heat current. Hence, we can conclude that $\rvert J_Q\arrowvert$ is maximum when both $\chi_{L(R)}=-0.5$, irrespective of $\xi_{L(R)}$, which is supported by numerical simulations [Fig.~\ref{fig derivative}(b,c)] and also follows from analytically derived condition $f_{L(R)}=1$ [Cf. Eq.~\eqref{A19}] for maximum heat current [Fig.~\ref{fig:3}:Inset]. Now, putting these conditions in Eq.~\eqref{J2}, we can evaluate the exact analytical expression of the maximum heat current as
    \begin{widetext}
    	\begin{equation}\begin{split}
    &|J_Q|_{max}\\
    &=\frac{U(\gamma_L+\gamma_R)}{\gamma_L\gamma_R}\bigg\rvert\frac{\sinh {(\frac{1}{2\xi_L}-\frac{1}{2\xi_R})}}{2[1+\cosh(\frac{1}{2\xi_R})+\cosh(\frac{1}{2\xi_L})]+\cosh{(\frac{1}{2\xi_L}-\frac{1}{2\xi_R})}+\cosh{(\frac{1}{2\xi_R}-\frac{1}{2\xi_L})}}\bigg\arrowvert \\
    &\equiv U|\Gamma|_{max}.
    \end{split}
    \end{equation}
    \end{widetext} 
    Several remarks are now in order: 
    \begin{itemize}
    	\item Maximum heat current is obtained at $\chi_{L(R)}=-0.5$ and the magnitude only depends on the temperature of two leads, their tunnelling rates, and the Coulomb interaction between the dots. Since, $-0.5$ is the mean of $0$ and $-1$ which are transition points of $f_{L(R)}^1$ and $f_{L(R)}^2$ respectively (point $B$ in Fig.~\ref{fig:3}:main) and this mean is also independent of other controlling parameters, we term this number ``$-0.5$'' as the ``\textit{universal magic mean}''. Significance of the point $B$ lies on the fact that at the magic mean $\chi_{L(R)}=-0.5$, chemical potential of left (right) lead becomes exactly equal to mean of the transition energies $\omega_{21(31)}$ and $\omega_{43(42)}$ driven by that bath:
    	\begin{equation}\begin{split}
    	&\chi_{L(R)}=-0.5\\
    	\Rightarrow \quad &\varepsilon_{L(R)}-\mu_{L(R)}=-0.5 U\\
    	\Rightarrow \quad &\mu_{L(R)}=\frac{2\varepsilon_{L(R)}+U}{2}\\
    	\Rightarrow \quad &\mu_{L}=\frac{\omega_{43}+\omega_{21}}{2};\quad \mu_{R}=\frac{\omega_{42}+\omega_{31}}{2}.
    	\end{split}
    	\end{equation}
    	and therefore provides maximum control over $L(R)$ bath to guide both the absorption and decay simultaneously at maximum $\Gamma$, resulting maximum $\rvert J_Q\arrowvert$.

    	\item With the increase of $\xi_{L(R)}$, $|J_Q|$ also spreads out, keeping maxima point fixed at $\chi_{L(R)}=-0.5$. This precisely indicates that maximum heat current will invariably be obtained at $\chi_{L(R)}=-0.5$, irrespective of all $\xi_{L,R}$ [Fig.~\ref{fig derivative}(b,c)]. 3D plots along $\chi_L$ are squeezed for smaller values of $\xi_{L(R)}$[Fig.~\ref{fig derivative}(b)] (Assuming $\xi_L<\xi_R$ as in Fig.~\ref{fig:App1}) and if we increase $\xi_L(\xi_R)$, it gets expanded along both positive and negative $\chi_{L(R)}$, leaving out point of maxima intact at $\chi_{L(R)}=-0.5$ [Fig.~\ref{fig derivative}(c)]. This fundamental feature of the ``magic mean'' makes it truly \textit{universal}.

    \end{itemize}
    
	\section{Efficient Heat current modulator}\label{HCM}
	
	\par
\begin{figure*}[t]
		\centering
		\includegraphics[width=18cm,height=5cm]{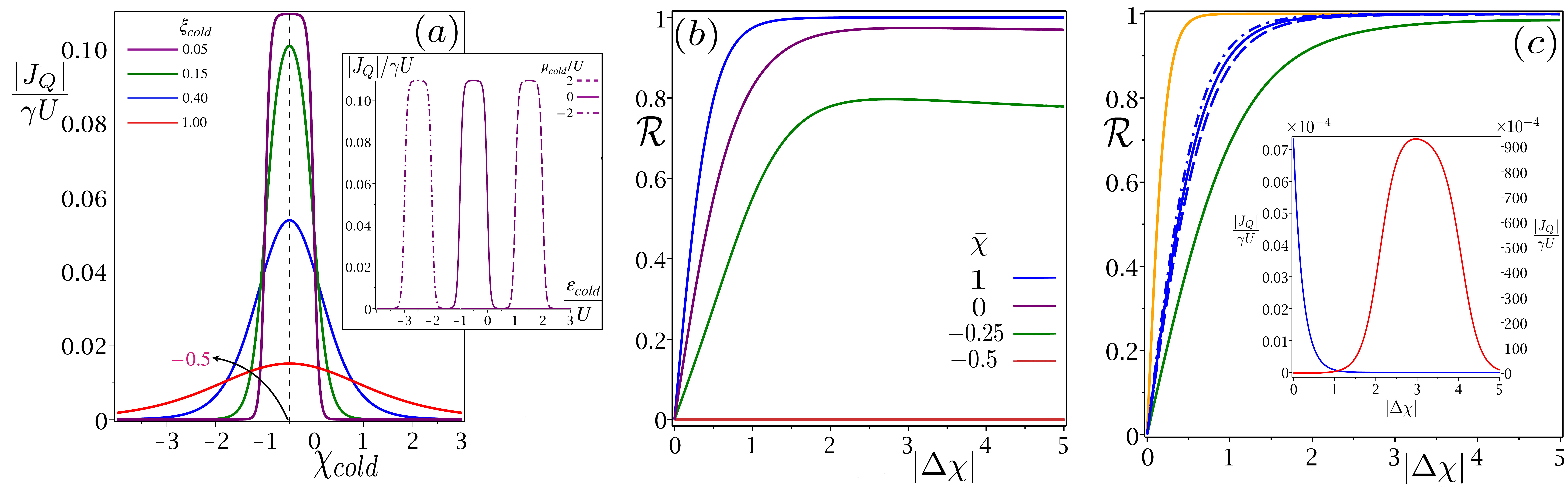}
		\caption{(a) Main: Thermal switching effect is obtained with the variation of scaled heat current $|J_Q|/\gamma U$ w.r.t $\chi_{cold}$ and gets more prominent with $\xi_{cold}\ll1$ as Domain ON and OFF become more precise; Inset: Expressing switching effect in terms of controllable parameter $\varepsilon_{L(R)}$. 
			(b) No rectification ($\mathcal{R}=0$) is obtained irrespective of the value of $\Delta \xi$, if $\bar{\chi}=-0.5$ or $\Delta{\chi}=0$ (red line). Partial (purple and green lines) and complete (blue line) rectifications are obtained for nonzero $|\Delta \chi|$ and $\bar{\chi}$ deviating from the magic mean $-0.5$. (c) Main: $\xi_{cold}$ (orange and green lines) has stronger dependence on rectification than $\xi_{hot}$ (blue dashed and dashed-dot lines) for an equal change in $|\Delta \xi|$, where solid blue line corresponds to the blue curve of Fig.\ref{fig:4}b; Inset: Magnitude of heat current decreases by an amount $10^{-6}$ on reversing the temperature gradient for the blue curve in Fig.\ref{fig:4}b.}
		\label{fig:4}
	\end{figure*}

 With the increase of $\xi_{L(R)}$, $\mathcal{D}_{1[5]}$ gets narrower while $\mathcal{D}_{2[4]}$ spreads out, without affecting $\mathcal{D}_3$ [Fig.~\ref{fig:3}]. As a consequence, $\arrowvert J_Q\rvert$ spreads out with $\xi_{L(R)}$, keeping point of maxima fixed at $\chi_{L(R)}=-0.5$. Now if $\xi_{L(R)} \ll 1$, then the span of $\mathcal{D}_{2[4]}$ is very small compared to $\mathcal{D}_{1[5]}$ while domain $\mathcal{D}_3$ always remains in between $-1 \leq \chi_{L(R)} \leq 0$. So, there are effectively two domains: (I) Domain ON, where $\arrowvert J_Q\rvert\neq0$ and (II) Domain OFF, where $\arrowvert J_Q\rvert=0$ and switching effect gets more prominent if $\xi_{L(R)} \ll 1$. Thus we can operate our model as an efficient thermal switch to on-off heat current just by shifting the domain from ON to OFF through the change of $\chi_{L(R)}$ or in turn controllable experimental parameter $\varepsilon_{L(R)}$ [Fig.~\ref{fig:4}a]. This is the basic underlying principle behind switching effect  for all coulomb coupled fermionic thermal diodes. It becomes more prominent for smaller cold bath temperature $\xi=k_BT/U$ and can be achieved more easily by varying $U$, instead of lowering $T$ to smaller value. 
	
	\par Finally, the model operates as an efficient thermal diode with rectification factor ($\mathcal R$) approaching $1$, even in presence of low temperature gradient. This produces a clear advantage over all previously proposed models with $\mathcal{R}$ factor defined as~\cite{PRB2021,roberts2011Sciencedirect,segak2005PRL,wu2009PRL,PRLQTT2016},
	\begin{equation}\label{Rec}
	\mathcal{R}(\Delta\xi)=\frac{\arrowvert J_Q^R(\Delta \xi)-J_Q^L(-\Delta \xi)\arrowvert}{\arrowvert J_Q^R(\Delta \xi)+ J_Q^L(-\Delta \xi)\lvert},
	\end{equation}
	where $\Delta \xi = \xi_R-\xi_L$ is identified as the temperature gradient, for $\xi_R>\xi_L$. With $\Delta\xi>0$, i.e. $\xi_{R(L)} \equiv \xi_{hot(cold)}$, heat current flows from $R$ to $L$ fulfilling $J_Q^R>0$ and if we exchange temperatures of the bath, then temperature gradient becomes $-\Delta\xi$ or $\xi_{R(L)} \equiv \xi_{cold(hot)}$ and consequently heat flows from $L$ to $R$ satisfying $J_Q^L>0$. Now if the heat current vanishes on reversing the temperature gradient, i.e., $\arrowvert J_Q\rvert$ is finite in one direction but null in the other, then complete rectification is achieved with $\mathcal R \rightarrow 1$, irrespective of the value of $\Delta \xi$, whereas $\mathcal R \rightarrow 0$ corresponds to no rectification i.e. no change in $\arrowvert J_Q\rvert$ upon inverting the temperature gradient. So, current asymmetry in two directions is primary criteria behind positive rectification: (i) It is clear from Fig.~\ref{fig:3} that for a given $\Delta \xi$, $\arrowvert J_Q\rvert$ depends on $|\Delta \chi|$. If $\chi_{hot}=\chi_{cold}$, or $\Delta\chi=\chi_{hot}-\chi_{cold}=0$, heat currents will be symmetric in both ways, therefore $\mathcal{R}=0$. (ii) From Fig.~\ref{fig:4}a, we find the variation of $\arrowvert J_Q\rvert$ w.r.t $\chi_{\alpha}$, is completely symmetric about the magic mean $-0.5$, i.e., $\rvert J_Q (\chi_{\alpha})\lvert=\rvert J_Q (-1-\chi_{\alpha})\arrowvert$. As a result, following Eq.~\eqref{Rec} and Fig.~\ref{fig:4}a, we finds if $\chi_{hot}=-1-\chi_{cold}$ or $\bar{\chi}=\frac{1}{2} (\chi_{hot}+\chi_{cold})=-0.5$, then $J_Q^R(\Delta \xi)= J_Q^L(-\Delta \xi)$, yielding $\mathcal{R}=0$. Thus, $\mathcal{R}$ can be zero either (i) the difference between effective tunnelling barriers $\Delta\chi=0$, or (ii) mean effective tunnelling barrier is equal to magic mean ($\bar{\chi}=-0.5$); else rectification occurs.

	Once $|\Delta\chi|\neq0$ and $\bar{\chi}\neq-0.5$, effective tunnelling barriers $\chi_{L(R)}$ make dissimilar effects on $|J_Q|$ upon reversing the temperature gradient and consequently $\mathcal{R}$ becomes nonzero. For a fixed value of $|\Delta \xi|$, $\mathcal{R}$ increases as $\Delta\chi$ shifts from zero and approaches one as $\bar{\chi}$ significantly deviates from the magic mean [Fig.~\ref{fig:4}b]. The reason behind this, for smaller $|\Delta\chi|$, the effect of tunnelling barriers on $|J_Q|$ are comparable on reversing the temperature gradient, yielding partial rectification ($0<\mathcal{R}<1$). But, with the increase in $|\Delta\chi|$, the effect of $\chi_{L(R)}$ are no longer compatible and it creates larger heat current asymmetry between the two directions, leading to complete rectification behaviour with $\mathcal{R} \rightarrow 1$. As heat current is symmetric about $\bar{\chi}=-0.5$, we only consider positive variation of $\bar{\chi}$ in Fig.\ref{fig:4}b. Moreover, variation of $\mathcal{R}$ with $|\Delta\chi|$ depends on the value of $|\Delta\xi|$, larger $\Delta\xi$ implies better rectification. As $\xi_{cold}$ is less than $\xi_{hot}$ and the transition cycle is always initiated by the cold bath, heat current executes stronger dependence on $\xi_{cold}$ than $\xi_{hot}$ for a given change in $|\Delta\xi|$. Thus the variation of $\mathcal{R}$ with $|\Delta\chi|$ yields appreciable change when a given $|\Delta\xi|$ is altered due to change in $\xi_{cold}$ than $\xi_{hot}$ [Fig.\ref{fig:4}c: Main]. So, the complete rectification is more favourable when $\xi_{cold}\ll1$ [Fig.~\ref{fig:4}a]. We may have complete rectification even without considering absolute temperature of the cold bath close to zero, as we can vary $\xi_{cold}=k_BT/U$ by changing $U$ and keeping $T$ finite. Finally, Fig.~\ref{fig:4}c[Inset] shows that heat currents become $\sim 10^{-6}$ times smaller on reversing temperature gradient which corresponds to $\mathcal{R}\approx1$ curve in Fig.\ref{fig:4}b.
	
	\section{Conclusion}\label{Con}
	
	\par To conclude, the present model can be implemented as an efficient thermal switch as well as thermal rectifier to modulate heat current close to ideal rectification, even at arbitrarily low temperature differences between the heat reservoirs. Independent of the details of the system, the heat flow and heat rectification are characterized by a small set of universal parameters. The position of maximum heat current at the \textit{magic mean} ``$-0.5$'' in terms of dimensionless physical parameters, is the major findings of the domain analysis scheme presented here. The magic mean is robust and \textit{universal} in the sense that it is invariant w.r.t the variation of any other system or bath parameters and truly reflects the impact of chemical potential to decide the magnitudes of heat current. Present protocol is unique to fermionic systems and can be applied to more complicated three or multi-terminal fermionic devices. The straightforward generalisation of the present scheme to multi-terminal set-up would be computing the magic mean associated with the multiple fermionic reservoirs.  
	The advantage of using quantum dot systems is that they have discrete energy levels with strong on-site Coulomb interaction, and can be simultaneously tunnel coupled to their respective reservoirs. Their discrete energy levels provides energy-selective transport and can be tunned via the application of the external gate voltages. In view of recent experimental advances in Coulomb coupled quantum-dot systems~\cite{thierschmann2015NatureNano,hartmann2015PRL,thierschmann2015NJP,pfeffer2015PRApp}, our findings will have important implications in designing novel thermal devices and opening up potential applications in controlling thermal current at nanoscales. 
	
	\section*{Acknowledgements}
	This research was funded by the Initiation grant of IITK (Grant No. IITK/CHM/2018513) and SRG, SERB (grant no. SRG/2019/000289) India. S.G. and N.G. are thankful to CSIR for the fellowship. S.G. is also grateful to the Ministry of Education, Government of India for the Prime Minister Research Fellowship (PMRF).
	
	\newpage
	\onecolumngrid
	\appendix

	\section{Derivation of the Lindblad Master equation}\label{Appendix-A}
	
	\par First we derive the master equation for our composite quantum dot system which is coupled with fermionic reservoirs. Let us start with the total interaction Hamiltonian given by
	\begin{equation}\label{H_T}
	H_T=\sum_{\alpha=L,R} H_T^\alpha = \sum_{\alpha=L,R} \sum_k (t_{\alpha k} c_{\alpha k}^\dagger d_\alpha+t_{\alpha k}^* d_\alpha^\dagger c_{\alpha k}).
	\end{equation}
	It should be noted that one can not write the fermionic interaction Hamiltonian using a tensor product representation, since operators involved in the tensor product commute by construction while fermionic operators anticommute. To formally derive the master equation using tensor product representation, one needs more formal approach in terms of Jordon-Wigner transformation. For details we refer to Ref \cite{schaller2009PRB,zedler2009PRB}. In the present case, starting from Eq.~\eqref{H_T}, we can derive the von-Neumann equation for the total density operator of the composite system $\rho_T(t)$,  as
	\begin{equation}
	\frac{d}{dt}\rho_T(t)=-\frac{i}{\hbar}[H_T(t),\rho_T(t)].
	\end{equation}
	Integrating the above equation following Ref.~\cite{zedler2009PRB}, one obtains
	\begin{equation}\begin{split}
	\frac{\partial}{\partial t}\rho_D(t)=\frac{1}{(i\hbar)^2}\int_0^t dt^\prime {\rm Tr}_{L,R}[H_T(t),[H_T(t^\prime),\rho_T(t^\prime)]],
	\end{split}
	\end{equation}
	where we denote ${\rm Tr}_{L,R}\{ \rho_T(t)\}=\rho_D(t)$ as the reduced density operator for the system and also assumed that ${\rm Tr}_{L,R}[H_T(t),\rho_T(0)]=0$. Here ${\rm Tr}_{L,R}$ refers to the trace over each bath degrees of freedom. The reduced dynamics of the system in the weak sequential tunnelling limit can then be written as ~\cite{PRB2018REC,PRB2021QTT,carmichael2002book,
		Book1bre2002}
	\begin{equation}\begin{split}\label{rho_D(t)}
	\dot\rho_D(t)=\frac{1}{(i\hbar)^2}\int_0^\infty dt^\prime {\rm Tr}_{L,R}[H_T(t),[H_T(t-t^\prime),\rho_D(t)\otimes\rho_L\otimes\rho_R]],
	\end{split}
	\end{equation}
	where we substitute $\rho_T(t)=\rho_D(t)\otimes\rho_L\otimes\rho_R$. Since the bath operator obeys ${\rm Tr}_\alpha\{c_\alpha(t)\rho_\alpha\}=0={\rm Tr}_\alpha\{c^{\dag}_\alpha(t)\rho_\alpha\}$ for $\alpha=L,R$, we obtain ~\cite{Book1bre2002,PRB2020}
	\begin{equation}
	{\rm Tr}_{L,R}\{[H_T^\alpha(t),[H_T^\beta(t-t^\prime),\rho_D(t)\otimes\rho_L\otimes\rho_R]]\}=0 \quad \quad \alpha\neq\beta;\alpha,\beta=L,R.
	\end{equation}
	As a result Eq.~\eqref{rho_D(t)} simplifies to
	\begin{equation}
	\dot\rho_D(t)=\frac{1}{(i\hbar)^2}\sum_{\alpha=L,R}\left\{\int_0^\infty dt^\prime\ {\rm Tr}_{L,R}[H_T^\alpha(t),[H_T^\alpha(t-t^\prime),\rho_D(t)\otimes\rho_L\otimes\rho_R]]\right\}.
	\end{equation}
	Following the standard procedure of \cite{zedler2009PRB,ghosh2012PRE}, one can then derive the master equation
	\begin{equation}
	\dot\rho_D(t)=\mathcal{L}_L[\rho_D(t)]+\mathcal{L}_R[\rho_D(t)],
	\end{equation}
	where the Lindblad operators $\mathcal{L}_{\alpha}[\rho]$ are given by
	\begin{eqnarray}\label{Lindblad}
	\mathcal{L}_{\alpha}[\rho_D(t)]&\equiv&\mathcal{L}_{\alpha}[\rho]=\sum_{\omega_{\alpha}>0} \mathcal{G}_{{\alpha}}(\omega_{\alpha})\left[d_{\alpha}^\dagger(\omega_{\alpha})\rho d_{\alpha}(\omega_{\alpha})-\frac{1}{2}{\{\rho}, d_{\alpha}(\omega_{\alpha})d_{\alpha}^\dagger(\omega_{\alpha})\}\right] \nonumber\\
	&+&\mathcal{G}_{{\alpha}}(-\omega_{\alpha})
	\left[d_{\alpha}(\omega_{\alpha})\ \rho\ d_{\alpha}^\dagger(\omega_{\alpha})-\frac{1}{2}{\{\rho}, d_{\alpha}^\dagger(\omega_{\alpha})d_{\alpha}(\omega_{\alpha})\}\right].\nonumber\\
	\end{eqnarray}
	The operator $d_{\alpha}(\omega_{\alpha})$ assumes the form of $|i\rangle\langle j|$ ($i \neq j;\; i,j = 1,2,3,4$) and causes the transition driven by left (right) reservoir with positive energy $\omega_{\alpha}$ such that $\omega_L=\omega_{21},\; \omega_{43}$ and $\omega_R=\omega_{31},\; \omega_{42}$. In above Eq.\eqref{Lindblad} the temperature dependent bath auto-correlations functions are given by~\cite{zedler2009PRB,schaller2009PRB,ghosh2012PRE},
	\begin{eqnarray}
	\mathcal{G}_{\alpha}(\omega_{\alpha})=& \frac{\gamma_{\alpha}}{2} f_\alpha(\omega_\alpha); \quad
	\mathcal{G}_{\alpha}(-\omega_{\alpha})=&\frac{\gamma_{\alpha}}{2} (1-f_\alpha(\omega_\alpha)).
	\end{eqnarray}	
	So, using the above relations in Eq.~\eqref{Lindblad}, we obtain the final expression of the Lindbladian operator as
	\begin{eqnarray}\label{Lindblad2}
	\mathcal{L}_{\alpha}[\rho]&=&\sum_{\omega_{\alpha}>0} \frac{\gamma_{\alpha}}{2} f_\alpha(\omega_\alpha)\left[d_{\alpha}^\dagger(\omega_{\alpha})\rho d_{\alpha}(\omega_{\alpha})-\frac{1}{2}{\{\rho}, d_{\alpha}(\omega_{\alpha})d_{\alpha}^\dagger(\omega_{\alpha})\}\right] \nonumber\\
	&+&\frac{\gamma_{\alpha}}{2} (1-f_\alpha(\omega_\alpha))
	\left[d_{\alpha}(\omega_{\alpha})\ \rho\ d_{\alpha}^\dagger(\omega_{\alpha})-\frac{1}{2}{\{\rho}, d_{\alpha}^\dagger(\omega_{\alpha})d_{\alpha}(\omega_{\alpha})\}\right].\nonumber\\
	\end{eqnarray}
	%\bibliographystyle{apsrev4-1}
	%\bibliography{Document2022}

\begin{thebibliography}{55}%
		\makeatletter
		\providecommand \@ifxundefined [1]{%
			\@ifx{#1\undefined}
		}%
		\providecommand \@ifnum [1]{%
			\ifnum #1\expandafter \@firstoftwo
			\else \expandafter \@secondoftwo
			\fi
		}%
		\providecommand \@ifx [1]{%
			\ifx #1\expandafter \@firstoftwo
			\else \expandafter \@secondoftwo
			\fi
		}%
		\providecommand \natexlab [1]{#1}%
		\providecommand \enquote  [1]{``#1''}%
		\providecommand \bibnamefont  [1]{#1}%
		\providecommand \bibfnamefont [1]{#1}%
		\providecommand \citenamefont [1]{#1}%
		\providecommand \href@noop [0]{\@secondoftwo}%
		\providecommand \href [0]{\begingroup \@sanitize@url \@href}%
		\providecommand \@href[1]{\@@startlink{#1}\@@href}%
		\providecommand \@@href[1]{\endgroup#1\@@endlink}%
		\providecommand \@sanitize@url [0]{\catcode `\\12\catcode `\$12\catcode
			`\&12\catcode `\#12\catcode `\^12\catcode `\_12\catcode `\%12\relax}%
		\providecommand \@@startlink[1]{}%
		\providecommand \@@endlink[0]{}%
		\providecommand \url  [0]{\begingroup\@sanitize@url \@url }%
		\providecommand \@url [1]{\endgroup\@href {#1}{\urlprefix }}%
		\providecommand \urlprefix  [0]{URL }%
		\providecommand \Eprint [0]{\href }%
		\providecommand \doibase [0]{http://dx.doi.org/}%
		\providecommand \selectlanguage [0]{\@gobble}%
		\providecommand \bibinfo  [0]{\@secondoftwo}%
		\providecommand \bibfield  [0]{\@secondoftwo}%
		\providecommand \translation [1]{[#1]}%
		\providecommand \BibitemOpen [0]{}%
		\providecommand \bibitemStop [0]{}%
		\providecommand \bibitemNoStop [0]{.\EOS\space}%
		\providecommand \EOS [0]{\spacefactor3000\relax}%
		\providecommand \BibitemShut  [1]{\csname bibitem#1\endcsname}%
		\let\auto@bib@innerbib\@empty
		%</preamble>
		\bibitem [{\citenamefont {Giazotto}\ \emph {et~al.}(2006)\citenamefont
			{Giazotto}, \citenamefont {Heikkil\"a}, \citenamefont {Luukanen},
			\citenamefont {Savin},\ and\ \citenamefont {Pekola}}]{giazotto2006RMP}%
		\BibitemOpen
		\bibfield  {author} {\bibinfo {author} {\bibfnamefont {F.}~\bibnamefont
				{Giazotto}}, \bibinfo {author} {\bibfnamefont {T.~T.}\ \bibnamefont
				{Heikkil\"a}}, \bibinfo {author} {\bibfnamefont {A.}~\bibnamefont
				{Luukanen}}, \bibinfo {author} {\bibfnamefont {A.~M.}\ \bibnamefont {Savin}},
			\ and\ \bibinfo {author} {\bibfnamefont {J.~P.}\ \bibnamefont {Pekola}},\
		}\href {\doibase 10.1103/RevModPhys.78.217} {\bibfield  {journal} {\bibinfo
				{journal} {Rev. Mod. Phys.}\ }\textbf {\bibinfo {volume} {78}},\ \bibinfo
			{pages} {217} (\bibinfo {year} {2006})}\BibitemShut {NoStop}%
		\bibitem [{\citenamefont {Pekola}(2015)}]{pekola2015Naturephysics}%
		\BibitemOpen
		\bibfield  {author} {\bibinfo {author} {\bibfnamefont {J.~P.}\ \bibnamefont
				{Pekola}},\ }\href {\doibase 10.1038/nphys3169} {\bibfield  {journal}
			{\bibinfo  {journal} {Nature Physics}\ }\textbf {\bibinfo {volume} {11}},\
			\bibinfo {pages} {118} (\bibinfo {year} {2015})}\BibitemShut {NoStop}%
		\bibitem [{\citenamefont {Benenti}\ \emph {et~al.}(2017)\citenamefont
			{Benenti}, \citenamefont {Casati}, \citenamefont {Saito},\ and\ \citenamefont
			{Whitney}}]{beneti2017Physicsreport}%
		\BibitemOpen
		\bibfield  {author} {\bibinfo {author} {\bibfnamefont {G.}~\bibnamefont
				{Benenti}}, \bibinfo {author} {\bibfnamefont {G.}~\bibnamefont {Casati}},
			\bibinfo {author} {\bibfnamefont {K.}~\bibnamefont {Saito}}, \ and\ \bibinfo
			{author} {\bibfnamefont {R.}~\bibnamefont {Whitney}},\ }\href {\doibase
			https://doi.org/10.1016/j.physrep.2017.05.008} {\bibfield  {journal}
			{\bibinfo  {journal} {Physics Reports}\ }\textbf {\bibinfo {volume} {694}},\
			\bibinfo {pages} {1} (\bibinfo {year} {2017})}\BibitemShut {NoStop}%
		\bibitem [{\citenamefont {Roberts}\ and\ \citenamefont
			{Walker}(2011)}]{roberts2011Sciencedirect}%
		\BibitemOpen
		\bibfield  {author} {\bibinfo {author} {\bibfnamefont {N.}~\bibnamefont
				{Roberts}}\ and\ \bibinfo {author} {\bibfnamefont {D.}~\bibnamefont
				{Walker}},\ }\href {\doibase
			https://doi.org/10.1016/j.ijthermalsci.2010.12.004} {\bibfield  {journal}
			{\bibinfo  {journal} {International Journal of Thermal Sciences}\ }\textbf
			{\bibinfo {volume} {50}},\ \bibinfo {pages} {648} (\bibinfo {year}
			{2011})}\BibitemShut {NoStop}%
		\bibitem [{\citenamefont {Li}\ \emph {et~al.}(2012)\citenamefont {Li},
			\citenamefont {Ren}, \citenamefont {Wang}, \citenamefont {Zhang},
			\citenamefont {H\"anggi},\ and\ \citenamefont {Li}}]{li2012RMP}%
		\BibitemOpen
		\bibfield  {author} {\bibinfo {author} {\bibfnamefont {N.}~\bibnamefont
				{Li}}, \bibinfo {author} {\bibfnamefont {J.}~\bibnamefont {Ren}}, \bibinfo
			{author} {\bibfnamefont {L.}~\bibnamefont {Wang}}, \bibinfo {author}
			{\bibfnamefont {G.}~\bibnamefont {Zhang}}, \bibinfo {author} {\bibfnamefont
				{P.}~\bibnamefont {H\"anggi}}, \ and\ \bibinfo {author} {\bibfnamefont
				{B.}~\bibnamefont {Li}},\ }\href {\doibase 10.1103/RevModPhys.84.1045}
		{\bibfield  {journal} {\bibinfo  {journal} {Rev. Mod. Phys.}\ }\textbf
			{\bibinfo {volume} {84}},\ \bibinfo {pages} {1045} (\bibinfo {year}
			{2012})}\BibitemShut {NoStop}%
		\bibitem [{\citenamefont {Naseem}\ \emph {et~al.}(2020)\citenamefont {Naseem},
			\citenamefont {Misra}, \citenamefont {M\"ustecaplio\ifmmode~\breve{g}\else
				\u{g}\fi{}lu},\ and\ \citenamefont {Kurizki}}]{kurizki2020PRR}%
		\BibitemOpen
		\bibfield  {author} {\bibinfo {author} {\bibfnamefont {M.~T.}\ \bibnamefont
				{Naseem}}, \bibinfo {author} {\bibfnamefont {A.}~\bibnamefont {Misra}},
			\bibinfo {author} {\bibfnamefont {O.~E.}\ \bibnamefont
				{M\"ustecaplio\ifmmode~\breve{g}\else \u{g}\fi{}lu}}, \ and\ \bibinfo
			{author} {\bibfnamefont {G.}~\bibnamefont {Kurizki}},\ }\href {\doibase
			10.1103/PhysRevResearch.2.033285} {\bibfield  {journal} {\bibinfo  {journal}
				{Phys. Rev. Research}\ }\textbf {\bibinfo {volume} {2}},\ \bibinfo {pages}
			{033285} (\bibinfo {year} {2020})}\BibitemShut {NoStop}%
		\bibitem [{\citenamefont {Zeng}\ and\ \citenamefont
			{Wang}(2008)}]{zeng2008PRB}%
		\BibitemOpen
		\bibfield  {author} {\bibinfo {author} {\bibfnamefont {N.}~\bibnamefont
				{Zeng}}\ and\ \bibinfo {author} {\bibfnamefont {J.-S.}\ \bibnamefont
				{Wang}},\ }\href {\doibase 10.1103/PhysRevB.78.024305} {\bibfield  {journal}
			{\bibinfo  {journal} {Phys. Rev. B}\ }\textbf {\bibinfo {volume} {78}},\
			\bibinfo {pages} {024305} (\bibinfo {year} {2008})}\BibitemShut {NoStop}%
		\bibitem [{\citenamefont {Ruokola}\ \emph {et~al.}(2009)\citenamefont
			{Ruokola}, \citenamefont {Ojanen},\ and\ \citenamefont
			{Jauho}}]{ruokola2009PRB}%
		\BibitemOpen
		\bibfield  {author} {\bibinfo {author} {\bibfnamefont {T.}~\bibnamefont
				{Ruokola}}, \bibinfo {author} {\bibfnamefont {T.}~\bibnamefont {Ojanen}}, \
			and\ \bibinfo {author} {\bibfnamefont {A.-P.}\ \bibnamefont {Jauho}},\ }\href
		{\doibase 10.1103/PhysRevB.79.144306} {\bibfield  {journal} {\bibinfo
				{journal} {Phys. Rev. B}\ }\textbf {\bibinfo {volume} {79}},\ \bibinfo
			{pages} {144306} (\bibinfo {year} {2009})}\BibitemShut {NoStop}%
		\bibitem [{\citenamefont {Kuo}\ and\ \citenamefont {Chang}(2010)}]{kuo2010PRB}%
		\BibitemOpen
		\bibfield  {author} {\bibinfo {author} {\bibfnamefont {D.~M.-T.}\
				\bibnamefont {Kuo}}\ and\ \bibinfo {author} {\bibfnamefont {Y.-c.}\
				\bibnamefont {Chang}},\ }\href {\doibase 10.1103/PhysRevB.81.205321}
		{\bibfield  {journal} {\bibinfo  {journal} {Phys. Rev. B}\ }\textbf {\bibinfo
				{volume} {81}},\ \bibinfo {pages} {205321} (\bibinfo {year}
			{2010})}\BibitemShut {NoStop}%
		\bibitem [{\citenamefont {Liu}\ \emph {et~al.}(2014)\citenamefont {Liu},
			\citenamefont {Zhou}, \citenamefont {Tang},\ and\ \citenamefont
			{Chen}}]{liu2014APL}%
		\BibitemOpen
		\bibfield  {author} {\bibinfo {author} {\bibfnamefont {Y.-Y.}\ \bibnamefont
				{Liu}}, \bibinfo {author} {\bibfnamefont {W.-X.}\ \bibnamefont {Zhou}},
			\bibinfo {author} {\bibfnamefont {L.-M.}\ \bibnamefont {Tang}}, \ and\
			\bibinfo {author} {\bibfnamefont {K.-Q.}\ \bibnamefont {Chen}},\ }\href
		{\doibase 10.1063/1.4902427} {\bibfield  {journal} {\bibinfo  {journal}
				{Applied Physics Letters}\ }\textbf {\bibinfo {volume} {105}},\ \bibinfo
			{pages} {203111} (\bibinfo {year} {2014})}\BibitemShut {NoStop}%
		\bibitem [{\citenamefont {S{\'{a}}nchez}\ \emph {et~al.}(2017)\citenamefont
			{S{\'{a}}nchez}, \citenamefont {Thierschmann},\ and\ \citenamefont
			{Molenkamp}}]{rafael2017NJP}%
		\BibitemOpen
		\bibfield  {author} {\bibinfo {author} {\bibfnamefont {R.}~\bibnamefont
				{S{\'{a}}nchez}}, \bibinfo {author} {\bibfnamefont {H.}~\bibnamefont
				{Thierschmann}}, \ and\ \bibinfo {author} {\bibfnamefont {L.~W.}\
				\bibnamefont {Molenkamp}},\ }\href {\doibase 10.1088/1367-2630/aa8b94}
		{\bibfield  {journal} {\bibinfo  {journal} {New Journal of Physics}\ }\textbf
			{\bibinfo {volume} {19}},\ \bibinfo {pages} {113040} (\bibinfo {year}
			{2017})}\BibitemShut {NoStop}%
		\bibitem [{\citenamefont {Landi}\ \emph {et~al.}(2021)\citenamefont {Landi},
			\citenamefont {Poletti},\ and\ \citenamefont {Schaller}}]{landi2021arxiv}%
		\BibitemOpen
		\bibfield  {author} {\bibinfo {author} {\bibfnamefont {G.~T.}\ \bibnamefont
				{Landi}}, \bibinfo {author} {\bibfnamefont {D.}~\bibnamefont {Poletti}}, \
			and\ \bibinfo {author} {\bibfnamefont {G.}~\bibnamefont {Schaller}},\ }\href
		{\doibase 10.48550/arxiv.2104.14350} {\  (\bibinfo {year} {2021}),\
			10.48550/arxiv.2104.14350}\BibitemShut {NoStop}%
		\bibitem [{\citenamefont {Ruokola}\ and\ \citenamefont
			{Ojanen}(2011)}]{PRBDIODE2011}%
		\BibitemOpen
		\bibfield  {author} {\bibinfo {author} {\bibfnamefont {T.}~\bibnamefont
				{Ruokola}}\ and\ \bibinfo {author} {\bibfnamefont {T.}~\bibnamefont
				{Ojanen}},\ }\href {\doibase 10.1103/PhysRevB.83.241404} {\bibfield
			{journal} {\bibinfo  {journal} {Phys. Rev. B}\ }\textbf {\bibinfo {volume}
				{83}},\ \bibinfo {pages} {241404} (\bibinfo {year} {2011})}\BibitemShut
		{NoStop}%
		\bibitem [{\citenamefont {Terraneo}\ \emph {et~al.}(2002)\citenamefont
			{Terraneo}, \citenamefont {Peyrard},\ and\ \citenamefont
			{Casati}}]{terraneo2002PRL}%
		\BibitemOpen
		\bibfield  {author} {\bibinfo {author} {\bibfnamefont {M.}~\bibnamefont
				{Terraneo}}, \bibinfo {author} {\bibfnamefont {M.}~\bibnamefont {Peyrard}}, \
			and\ \bibinfo {author} {\bibfnamefont {G.}~\bibnamefont {Casati}},\ }\href
		{\doibase 10.1103/PhysRevLett.88.094302} {\bibfield  {journal} {\bibinfo
				{journal} {Phys. Rev. Lett.}\ }\textbf {\bibinfo {volume} {88}},\ \bibinfo
			{pages} {094302} (\bibinfo {year} {2002})}\BibitemShut {NoStop}%
		\bibitem [{\citenamefont {Li}\ \emph {et~al.}(2004)\citenamefont {Li},
			\citenamefont {Wang},\ and\ \citenamefont {Casati}}]{li2004PRL}%
		\BibitemOpen
		\bibfield  {author} {\bibinfo {author} {\bibfnamefont {B.}~\bibnamefont
				{Li}}, \bibinfo {author} {\bibfnamefont {L.}~\bibnamefont {Wang}}, \ and\
			\bibinfo {author} {\bibfnamefont {G.}~\bibnamefont {Casati}},\ }\href
		{\doibase 10.1103/PhysRevLett.93.184301} {\bibfield  {journal} {\bibinfo
				{journal} {Phys. Rev. Lett.}\ }\textbf {\bibinfo {volume} {93}},\ \bibinfo
			{pages} {184301} (\bibinfo {year} {2004})}\BibitemShut {NoStop}%
		\bibitem [{\citenamefont {Segal}\ and\ \citenamefont
			{Nitzan}(2005)}]{segak2005PRL}%
		\BibitemOpen
		\bibfield  {author} {\bibinfo {author} {\bibfnamefont {D.}~\bibnamefont
				{Segal}}\ and\ \bibinfo {author} {\bibfnamefont {A.}~\bibnamefont {Nitzan}},\
		}\href {\doibase 10.1103/PhysRevLett.94.034301} {\bibfield  {journal}
			{\bibinfo  {journal} {Phys. Rev. Lett.}\ }\textbf {\bibinfo {volume} {94}},\
			\bibinfo {pages} {034301} (\bibinfo {year} {2005})}\BibitemShut {NoStop}%
		\bibitem [{\citenamefont {Ojanen}(2009)}]{PRB2009}%
		\BibitemOpen
		\bibfield  {author} {\bibinfo {author} {\bibfnamefont {T.}~\bibnamefont
				{Ojanen}},\ }\href {\doibase 10.1103/PhysRevB.80.180301} {\bibfield
			{journal} {\bibinfo  {journal} {Phys. Rev. B}\ }\textbf {\bibinfo {volume}
				{80}},\ \bibinfo {pages} {180301} (\bibinfo {year} {2009})}\BibitemShut
		{NoStop}%
		\bibitem [{\citenamefont {Wu}\ and\ \citenamefont {Segal}(2009)}]{wu2009PRL}%
		\BibitemOpen
		\bibfield  {author} {\bibinfo {author} {\bibfnamefont {L.-A.}\ \bibnamefont
				{Wu}}\ and\ \bibinfo {author} {\bibfnamefont {D.}~\bibnamefont {Segal}},\
		}\href {\doibase 10.1103/PhysRevLett.102.095503} {\bibfield  {journal}
			{\bibinfo  {journal} {Phys. Rev. Lett.}\ }\textbf {\bibinfo {volume} {102}},\
			\bibinfo {pages} {095503} (\bibinfo {year} {2009})}\BibitemShut {NoStop}%
		\bibitem [{\citenamefont {Werlang}\ \emph {et~al.}(2014)\citenamefont
			{Werlang}, \citenamefont {Marchiori}, \citenamefont {Cornelio},\ and\
			\citenamefont {Valente}}]{PREREC2014}%
		\BibitemOpen
		\bibfield  {author} {\bibinfo {author} {\bibfnamefont {T.}~\bibnamefont
				{Werlang}}, \bibinfo {author} {\bibfnamefont {M.~A.}\ \bibnamefont
				{Marchiori}}, \bibinfo {author} {\bibfnamefont {M.~F.}\ \bibnamefont
				{Cornelio}}, \ and\ \bibinfo {author} {\bibfnamefont {D.}~\bibnamefont
				{Valente}},\ }\href {\doibase 10.1103/PhysRevE.89.062109} {\bibfield
			{journal} {\bibinfo  {journal} {Phys. Rev. E}\ }\textbf {\bibinfo {volume}
				{89}},\ \bibinfo {pages} {062109} (\bibinfo {year} {2014})}\BibitemShut
		{NoStop}%
		\bibitem [{\citenamefont {Mascarenhas}\ \emph {et~al.}(2014)\citenamefont
			{Mascarenhas}, \citenamefont {Gerace}, \citenamefont {Valente}, \citenamefont
			{Montangero}, \citenamefont {Auff{\`{e}}ves},\ and\ \citenamefont
			{Santos}}]{Mascarenhas2014IOP}%
		\BibitemOpen
		\bibfield  {author} {\bibinfo {author} {\bibfnamefont {E.}~\bibnamefont
				{Mascarenhas}}, \bibinfo {author} {\bibfnamefont {D.}~\bibnamefont {Gerace}},
			\bibinfo {author} {\bibfnamefont {D.}~\bibnamefont {Valente}}, \bibinfo
			{author} {\bibfnamefont {S.}~\bibnamefont {Montangero}}, \bibinfo {author}
			{\bibfnamefont {A.}~\bibnamefont {Auff{\`{e}}ves}}, \ and\ \bibinfo {author}
			{\bibfnamefont {M.~F.}\ \bibnamefont {Santos}},\ }\href {\doibase
			10.1209/0295-5075/106/54003} {\bibfield  {journal} {\bibinfo  {journal}
				{{EPL} (Europhysics Letters)}\ }\textbf {\bibinfo {volume} {106}},\ \bibinfo
			{pages} {54003} (\bibinfo {year} {2014})}\BibitemShut {NoStop}%
		\bibitem [{\citenamefont {Marcos-Vicioso}\ \emph {et~al.}(2018)\citenamefont
			{Marcos-Vicioso}, \citenamefont {L\'opez-Jurado}, \citenamefont
			{Ruiz-Garcia},\ and\ \citenamefont {S\'anchez}}]{PRB2018REC}%
		\BibitemOpen
		\bibfield  {author} {\bibinfo {author} {\bibfnamefont {A.}~\bibnamefont
				{Marcos-Vicioso}}, \bibinfo {author} {\bibfnamefont {C.}~\bibnamefont
				{L\'opez-Jurado}}, \bibinfo {author} {\bibfnamefont {M.}~\bibnamefont
				{Ruiz-Garcia}}, \ and\ \bibinfo {author} {\bibfnamefont {R.}~\bibnamefont
				{S\'anchez}},\ }\href {\doibase 10.1103/PhysRevB.98.035414} {\bibfield
			{journal} {\bibinfo  {journal} {Phys. Rev. B}\ }\textbf {\bibinfo {volume}
				{98}},\ \bibinfo {pages} {035414} (\bibinfo {year} {2018})}\BibitemShut
		{NoStop}%
		\bibitem [{\citenamefont {Bhandari}\ \emph {et~al.}(2021)\citenamefont
			{Bhandari}, \citenamefont {Erdman}, \citenamefont {Fazio}, \citenamefont
			{Paladino},\ and\ \citenamefont {Taddei}}]{PRB2021}%
		\BibitemOpen
		\bibfield  {author} {\bibinfo {author} {\bibfnamefont {B.}~\bibnamefont
				{Bhandari}}, \bibinfo {author} {\bibfnamefont {P.~A.}\ \bibnamefont
				{Erdman}}, \bibinfo {author} {\bibfnamefont {R.}~\bibnamefont {Fazio}},
			\bibinfo {author} {\bibfnamefont {E.}~\bibnamefont {Paladino}}, \ and\
			\bibinfo {author} {\bibfnamefont {F.}~\bibnamefont {Taddei}},\ }\href
		{\doibase 10.1103/PhysRevB.103.155434} {\bibfield  {journal} {\bibinfo
				{journal} {Phys. Rev. B}\ }\textbf {\bibinfo {volume} {103}},\ \bibinfo
			{pages} {155434} (\bibinfo {year} {2021})}\BibitemShut {NoStop}%
		\bibitem [{\citenamefont {Iorio}\ \emph {et~al.}(2021)\citenamefont {Iorio},
			\citenamefont {Strambini}, \citenamefont {Haack}, \citenamefont {Campisi},\
			and\ \citenamefont {Giazotto}}]{iorio2021PRAPP}%
		\BibitemOpen
		\bibfield  {author} {\bibinfo {author} {\bibfnamefont {A.}~\bibnamefont
				{Iorio}}, \bibinfo {author} {\bibfnamefont {E.}~\bibnamefont {Strambini}},
			\bibinfo {author} {\bibfnamefont {G.}~\bibnamefont {Haack}}, \bibinfo
			{author} {\bibfnamefont {M.}~\bibnamefont {Campisi}}, \ and\ \bibinfo
			{author} {\bibfnamefont {F.}~\bibnamefont {Giazotto}},\ }\href {\doibase
			10.1103/PhysRevApplied.15.054050} {\bibfield  {journal} {\bibinfo  {journal}
				{Phys. Rev. Applied}\ }\textbf {\bibinfo {volume} {15}},\ \bibinfo {pages}
			{054050} (\bibinfo {year} {2021})}\BibitemShut {NoStop}%
		\bibitem [{\citenamefont {Balachandran}\ \emph {et~al.}(2019)\citenamefont
			{Balachandran}, \citenamefont {Clark}, \citenamefont {Goold},\ and\
			\citenamefont {Poletti}}]{balachandran2019PRL}%
		\BibitemOpen
		\bibfield  {author} {\bibinfo {author} {\bibfnamefont {V.}~\bibnamefont
				{Balachandran}}, \bibinfo {author} {\bibfnamefont {S.~R.}\ \bibnamefont
				{Clark}}, \bibinfo {author} {\bibfnamefont {J.}~\bibnamefont {Goold}}, \ and\
			\bibinfo {author} {\bibfnamefont {D.}~\bibnamefont {Poletti}},\ }\href
		{\doibase 10.1103/PhysRevLett.123.020603} {\bibfield  {journal} {\bibinfo
				{journal} {Phys. Rev. Lett.}\ }\textbf {\bibinfo {volume} {123}},\ \bibinfo
			{pages} {020603} (\bibinfo {year} {2019})}\BibitemShut {NoStop}%
		\bibitem [{\citenamefont {Karg\ifmmode\imath\else\i\fi{}}\ \emph
			{et~al.}(2019)\citenamefont {Karg\ifmmode\imath\else\i\fi{}}, \citenamefont
			{Naseem}, \citenamefont {Opatrn\'y}, \citenamefont
			{M\"ustecapl\ifmmode\imath\else\i \fi{}o\ifmmode~\breve{g}\else
				\u{g}\fi{}lu},\ and\ \citenamefont {Kurizki}}]{kargi2019PRE}%
		\BibitemOpen
		\bibfield  {author} {\bibinfo {author} {\bibfnamefont {C.}~\bibnamefont
				{Karg\ifmmode\imath\else\i\fi{}}}, \bibinfo {author} {\bibfnamefont {M.~T.}\
				\bibnamefont {Naseem}}, \bibinfo {author} {\bibfnamefont {T.}~\bibnamefont
				{Opatrn\'y}}, \bibinfo {author} {\bibfnamefont {O.~E.}\ \bibnamefont
				{M\"ustecapl\ifmmode\imath\else\i \fi{}o\ifmmode~\breve{g}\else
					\u{g}\fi{}lu}}, \ and\ \bibinfo {author} {\bibfnamefont {G.}~\bibnamefont
				{Kurizki}},\ }\href {\doibase 10.1103/PhysRevE.99.042121} {\bibfield
			{journal} {\bibinfo  {journal} {Phys. Rev. E}\ }\textbf {\bibinfo {volume}
				{99}},\ \bibinfo {pages} {042121} (\bibinfo {year} {2019})}\BibitemShut
		{NoStop}%
		\bibitem [{\citenamefont {Ordonez-Miranda}\ \emph {et~al.}(2017)\citenamefont
			{Ordonez-Miranda}, \citenamefont {Ezzahri},\ and\ \citenamefont
			{Joulain}}]{PREREC2017}%
		\BibitemOpen
		\bibfield  {author} {\bibinfo {author} {\bibfnamefont {J.}~\bibnamefont
				{Ordonez-Miranda}}, \bibinfo {author} {\bibfnamefont {Y.}~\bibnamefont
				{Ezzahri}}, \ and\ \bibinfo {author} {\bibfnamefont {K.}~\bibnamefont
				{Joulain}},\ }\href {\doibase 10.1103/PhysRevE.95.022128} {\bibfield
			{journal} {\bibinfo  {journal} {Phys. Rev. E}\ }\textbf {\bibinfo {volume}
				{95}},\ \bibinfo {pages} {022128} (\bibinfo {year} {2017})}\BibitemShut
		{NoStop}%
		\bibitem [{\citenamefont {Aligia}\ \emph {et~al.}(2020)\citenamefont {Aligia},
			\citenamefont {Daroca}, \citenamefont {Arrachea},\ and\ \citenamefont
			{Roura-Bas}}]{PRB2020}%
		\BibitemOpen
		\bibfield  {author} {\bibinfo {author} {\bibfnamefont {A.~A.}\ \bibnamefont
				{Aligia}}, \bibinfo {author} {\bibfnamefont {D.~P.}\ \bibnamefont {Daroca}},
			\bibinfo {author} {\bibfnamefont {L.}~\bibnamefont {Arrachea}}, \ and\
			\bibinfo {author} {\bibfnamefont {P.}~\bibnamefont {Roura-Bas}},\ }\href
		{\doibase 10.1103/PhysRevB.101.075417} {\bibfield  {journal} {\bibinfo
				{journal} {Phys. Rev. B}\ }\textbf {\bibinfo {volume} {101}},\ \bibinfo
			{pages} {075417} (\bibinfo {year} {2020})}\BibitemShut {NoStop}%
		\bibitem [{\citenamefont {Upadhyay}\ \emph {et~al.}(2021)\citenamefont
			{Upadhyay}, \citenamefont {Naseem}, \citenamefont {Marathe},\ and\
			\citenamefont {M\"ustecapl\ifmmode \imath \else \i
				\fi{}o\ifmmode~\breve{g}\else \u{g}\fi{}lu}}]{upadhyay2021PRE}%
		\BibitemOpen
		\bibfield  {author} {\bibinfo {author} {\bibfnamefont {V.}~\bibnamefont
				{Upadhyay}}, \bibinfo {author} {\bibfnamefont {M.~T.}\ \bibnamefont
				{Naseem}}, \bibinfo {author} {\bibfnamefont {R.}~\bibnamefont {Marathe}}, \
			and\ \bibinfo {author} {\bibfnamefont {O.~E.}\ \bibnamefont
				{M\"ustecapl\ifmmode \imath \else \i \fi{}o\ifmmode~\breve{g}\else
					\u{g}\fi{}lu}},\ }\href {\doibase 10.1103/PhysRevE.104.054137} {\bibfield
			{journal} {\bibinfo  {journal} {Phys. Rev. E}\ }\textbf {\bibinfo {volume}
				{104}},\ \bibinfo {pages} {054137} (\bibinfo {year} {2021})}\BibitemShut
		{NoStop}%
		\bibitem [{\citenamefont {D{\'{\i}}az}\ and\ \citenamefont
			{S{\'{a}}nchez}(2021)}]{diaz_2021NJP}%
		\BibitemOpen
		\bibfield  {author} {\bibinfo {author} {\bibfnamefont {I.}~\bibnamefont
				{D{\'{\i}}az}}\ and\ \bibinfo {author} {\bibfnamefont {R.}~\bibnamefont
				{S{\'{a}}nchez}},\ }\href {\doibase 10.1088/1367-2630/ac4211} {\bibfield
			{journal} {\bibinfo  {journal} {New Journal of Physics}\ }\textbf {\bibinfo
				{volume} {23}},\ \bibinfo {pages} {125006} (\bibinfo {year}
			{2021})}\BibitemShut {NoStop}%
		\bibitem [{\citenamefont {Tesser}\ \emph {et~al.}(2022)\citenamefont {Tesser},
			\citenamefont {Bhandari}, \citenamefont {Erdman}, \citenamefont {Paladino},
			\citenamefont {Fazio},\ and\ \citenamefont {Taddei}}]{NJP2022}%
		\BibitemOpen
		\bibfield  {author} {\bibinfo {author} {\bibfnamefont {L.}~\bibnamefont
				{Tesser}}, \bibinfo {author} {\bibfnamefont {B.}~\bibnamefont {Bhandari}},
			\bibinfo {author} {\bibfnamefont {P.~A.}\ \bibnamefont {Erdman}}, \bibinfo
			{author} {\bibfnamefont {E.}~\bibnamefont {Paladino}}, \bibinfo {author}
			{\bibfnamefont {R.}~\bibnamefont {Fazio}}, \ and\ \bibinfo {author}
			{\bibfnamefont {F.}~\bibnamefont {Taddei}},\ }\href {\doibase
			10.1088/1367-2630/ac53b8} {\bibfield  {journal} {\bibinfo  {journal} {New
					Journal of Physics}\ }\textbf {\bibinfo {volume} {24}},\ \bibinfo {pages}
			{035001} (\bibinfo {year} {2022})}\BibitemShut {NoStop}%
		\bibitem [{\citenamefont {Zhang}\ \emph {et~al.}(2017)\citenamefont {Zhang},
			\citenamefont {Zhang}, \citenamefont {Ye}, \citenamefont {Lin},\ and\
			\citenamefont {Chen}}]{APL2017}%
		\BibitemOpen
		\bibfield  {author} {\bibinfo {author} {\bibfnamefont {Y.}~\bibnamefont
				{Zhang}}, \bibinfo {author} {\bibfnamefont {X.}~\bibnamefont {Zhang}},
			\bibinfo {author} {\bibfnamefont {Z.}~\bibnamefont {Ye}}, \bibinfo {author}
			{\bibfnamefont {G.}~\bibnamefont {Lin}}, \ and\ \bibinfo {author}
			{\bibfnamefont {J.}~\bibnamefont {Chen}},\ }\href {\doibase
			10.1063/1.4979977} {\bibfield  {journal} {\bibinfo  {journal} {Applied
					Physics Letters}\ }\textbf {\bibinfo {volume} {110}},\ \bibinfo {pages}
			{153501} (\bibinfo {year} {2017})}\BibitemShut {NoStop}%
		\bibitem [{\citenamefont {Zhang}\ \emph {et~al.}(2018)\citenamefont {Zhang},
			\citenamefont {Yang}, \citenamefont {Zhang}, \citenamefont {Lin},
			\citenamefont {Lin},\ and\ \citenamefont {Chen}}]{zhang2018EPL}%
		\BibitemOpen
		\bibfield  {author} {\bibinfo {author} {\bibfnamefont {Y.}~\bibnamefont
				{Zhang}}, \bibinfo {author} {\bibfnamefont {Z.}~\bibnamefont {Yang}},
			\bibinfo {author} {\bibfnamefont {X.}~\bibnamefont {Zhang}}, \bibinfo
			{author} {\bibfnamefont {B.}~\bibnamefont {Lin}}, \bibinfo {author}
			{\bibfnamefont {G.}~\bibnamefont {Lin}}, \ and\ \bibinfo {author}
			{\bibfnamefont {J.}~\bibnamefont {Chen}},\ }\href {\doibase
			10.1209/0295-5075/122/17002} {\bibfield  {journal} {\bibinfo  {journal}
				{{EPL} (Europhysics Letters)}\ }\textbf {\bibinfo {volume} {122}},\ \bibinfo
			{pages} {17002} (\bibinfo {year} {2018})}\BibitemShut {NoStop}%
		\bibitem [{\citenamefont {Ghosh}\ \emph {et~al.}(2012)\citenamefont {Ghosh},
			\citenamefont {Sinha},\ and\ \citenamefont {Ray}}]{ghosh2012PRE}%
		\BibitemOpen
		\bibfield  {author} {\bibinfo {author} {\bibfnamefont {A.}~\bibnamefont
				{Ghosh}}, \bibinfo {author} {\bibfnamefont {S.~S.}\ \bibnamefont {Sinha}}, \
			and\ \bibinfo {author} {\bibfnamefont {D.~S.}\ \bibnamefont {Ray}},\ }\href
		{\doibase 10.1103/PhysRevE.86.011138} {\bibfield  {journal} {\bibinfo
				{journal} {Phys. Rev. E}\ }\textbf {\bibinfo {volume} {86}},\ \bibinfo
			{pages} {011138} (\bibinfo {year} {2012})}\BibitemShut {NoStop}%
		\bibitem [{\citenamefont {Breuer}\ and\ \citenamefont
			{Petruccione}(2002)}]{Book1bre2002}%
		\BibitemOpen
		\bibfield  {author} {\bibinfo {author} {\bibfnamefont {H.-P.}\ \bibnamefont
				{Breuer}}\ and\ \bibinfo {author} {\bibfnamefont {F.}~\bibnamefont
				{Petruccione}},\ }\href@noop {} {\emph {\bibinfo {title} {The Theory of Open
					Quantum Systems}}}\ (\bibinfo  {publisher} {Oxford University Press,
			Oxford},\ \bibinfo {year} {2002})\BibitemShut {NoStop}%
		\bibitem [{\citenamefont {Carmichael}(2002)}]{carmichael2002book}%
		\BibitemOpen
		\bibfield  {author} {\bibinfo {author} {\bibfnamefont {H.~J.}\ \bibnamefont
				{Carmichael}},\ }\href@noop {} {\emph {\bibinfo {title} {Statistical Methods
					in Quantum Optics 1}}}\ (\bibinfo  {publisher} {Springer-Verlag},\ \bibinfo
		{address} {Berlin Heidelberg},\ \bibinfo {year} {2002})\BibitemShut {NoStop}%
		\bibitem [{\citenamefont {Joulain}\ \emph {et~al.}(2016)\citenamefont
			{Joulain}, \citenamefont {Drevillon}, \citenamefont {Ezzahri},\ and\
			\citenamefont {Ordonez-Miranda}}]{PRLQTT2016}%
		\BibitemOpen
		\bibfield  {author} {\bibinfo {author} {\bibfnamefont {K.}~\bibnamefont
				{Joulain}}, \bibinfo {author} {\bibfnamefont {J.}~\bibnamefont {Drevillon}},
			\bibinfo {author} {\bibfnamefont {Y.}~\bibnamefont {Ezzahri}}, \ and\
			\bibinfo {author} {\bibfnamefont {J.}~\bibnamefont {Ordonez-Miranda}},\
		}\href {\doibase 10.1103/PhysRevLett.116.200601} {\bibfield  {journal}
			{\bibinfo  {journal} {Phys. Rev. Lett.}\ }\textbf {\bibinfo {volume} {116}},\
			\bibinfo {pages} {200601} (\bibinfo {year} {2016})}\BibitemShut {NoStop}%
		\bibitem [{\citenamefont {Katz}\ and\ \citenamefont
			{Kosloff}(2016)}]{katz2016entropy}%
		\BibitemOpen
		\bibfield  {author} {\bibinfo {author} {\bibfnamefont {G.}~\bibnamefont
				{Katz}}\ and\ \bibinfo {author} {\bibfnamefont {R.}~\bibnamefont {Kosloff}},\
		}\href {https://www.mdpi.com/1099-4300/18/5/186} {\bibfield  {journal}
			{\bibinfo  {journal} {Entropy}\ }\textbf {\bibinfo {volume} {18}} (\bibinfo
			{year} {2016})}\BibitemShut {NoStop}%
		\bibitem [{\citenamefont {Jiang}\ \emph {et~al.}(2015)\citenamefont {Jiang},
			\citenamefont {Kulkarni}, \citenamefont {Segal},\ and\ \citenamefont
			{Imry}}]{jiang2015PRB}%
		\BibitemOpen
		\bibfield  {author} {\bibinfo {author} {\bibfnamefont {J.-H.}\ \bibnamefont
				{Jiang}}, \bibinfo {author} {\bibfnamefont {M.}~\bibnamefont {Kulkarni}},
			\bibinfo {author} {\bibfnamefont {D.}~\bibnamefont {Segal}}, \ and\ \bibinfo
			{author} {\bibfnamefont {Y.}~\bibnamefont {Imry}},\ }\href {\doibase
			10.1103/PhysRevB.92.045309} {\bibfield  {journal} {\bibinfo  {journal} {Phys.
					Rev. B}\ }\textbf {\bibinfo {volume} {92}},\ \bibinfo {pages} {045309}
			(\bibinfo {year} {2015})}\BibitemShut {NoStop}%
		\bibitem [{\citenamefont {Goury}\ and\ \citenamefont
			{Sánchez}(2019)}]{goury2019APL}%
		\BibitemOpen
		\bibfield  {author} {\bibinfo {author} {\bibfnamefont {D.}~\bibnamefont
				{Goury}}\ and\ \bibinfo {author} {\bibfnamefont {R.}~\bibnamefont
				{Sánchez}},\ }\href {\doibase 10.1063/1.5109100} {\bibfield  {journal}
			{\bibinfo  {journal} {Applied Physics Letters}\ }\textbf {\bibinfo {volume}
				{115}},\ \bibinfo {pages} {092601} (\bibinfo {year} {2019})}\BibitemShut
		{NoStop}%
		\bibitem [{\citenamefont {Liu}\ \emph {et~al.}(2022)\citenamefont {Liu},
			\citenamefont {Yu},\ and\ \citenamefont {Yu}}]{liu2022entropy}%
		\BibitemOpen
		\bibfield  {author} {\bibinfo {author} {\bibfnamefont {Y.-Q.}\ \bibnamefont
				{Liu}}, \bibinfo {author} {\bibfnamefont {D.-H.}\ \bibnamefont {Yu}}, \ and\
			\bibinfo {author} {\bibfnamefont {C.-S.}\ \bibnamefont {Yu}},\ }\href
		{https://www.mdpi.com/1099-4300/24/1/32} {\bibfield  {journal} {\bibinfo
				{journal} {Entropy}\ }\textbf {\bibinfo {volume} {24}} (\bibinfo {year}
			{2022})}\BibitemShut {NoStop}%
		\bibitem [{\citenamefont {Gupt}\ \emph {et~al.}(2022)\citenamefont {Gupt},
			\citenamefont {Bhattacharyya}, \citenamefont {Das}, \citenamefont {Datta},
			\citenamefont {Mukherjee},\ and\ \citenamefont {Ghosh}}]{gupt2022PRE}%
		\BibitemOpen
		\bibfield  {author} {\bibinfo {author} {\bibfnamefont {N.}~\bibnamefont
				{Gupt}}, \bibinfo {author} {\bibfnamefont {S.}~\bibnamefont {Bhattacharyya}},
			\bibinfo {author} {\bibfnamefont {B.}~\bibnamefont {Das}}, \bibinfo {author}
			{\bibfnamefont {S.}~\bibnamefont {Datta}}, \bibinfo {author} {\bibfnamefont
				{V.}~\bibnamefont {Mukherjee}}, \ and\ \bibinfo {author} {\bibfnamefont
				{A.}~\bibnamefont {Ghosh}},\ }\href {\doibase 10.1103/PhysRevE.106.024110}
		{\bibfield  {journal} {\bibinfo  {journal} {Phys. Rev. E}\ }\textbf {\bibinfo
				{volume} {106}},\ \bibinfo {pages} {024110} (\bibinfo {year}
			{2022})}\BibitemShut {NoStop}%
		\bibitem [{\citenamefont {Gelbwaser-Klimovsky}\ \emph
			{et~al.}(2015)\citenamefont {Gelbwaser-Klimovsky}, \citenamefont {Niedenzu},\
			and\ \citenamefont {Kurizki}}]{david2015Elsevier}%
		\BibitemOpen
		\bibfield  {author} {\bibinfo {author} {\bibfnamefont {D.}~\bibnamefont
				{Gelbwaser-Klimovsky}}, \bibinfo {author} {\bibfnamefont {W.}~\bibnamefont
				{Niedenzu}}, \ and\ \bibinfo {author} {\bibfnamefont {G.}~\bibnamefont
				{Kurizki}},\ }\href {\doibase https://doi.org/10.1016/bs.aamop.2015.07.002}
		{\ \bibinfo {series} {Advances In Atomic, Molecular, and Optical Physics},\
			\textbf {\bibinfo {volume} {64}},\ \bibinfo {pages} {329} (\bibinfo {year}
			{2015})}\BibitemShut {NoStop}%
		\bibitem [{\citenamefont {Spohn}(1978)}]{spohn1978JMP}%
		\BibitemOpen
		\bibfield  {author} {\bibinfo {author} {\bibfnamefont {H.}~\bibnamefont
				{Spohn}},\ }\href {\doibase 10.1063/1.523789} {\bibfield  {journal} {\bibinfo
				{journal} {Journal of Mathematical Physics}\ }\textbf {\bibinfo {volume}
				{19}},\ \bibinfo {pages} {1227} (\bibinfo {year} {1978})}\BibitemShut
		{NoStop}%
		\bibitem [{\citenamefont {Kosloff}(2013)}]{kosloff2013Entropy}%
		\BibitemOpen
		\bibfield  {author} {\bibinfo {author} {\bibfnamefont {R.}~\bibnamefont
				{Kosloff}},\ }\href {\doibase 10.3390/e15062100} {\bibfield  {journal}
			{\bibinfo  {journal} {Entropy}\ }\textbf {\bibinfo {volume} {15}},\ \bibinfo
			{pages} {2100} (\bibinfo {year} {2013})}\BibitemShut {NoStop}%
		\bibitem [{\citenamefont {Deffner}\ and\ \citenamefont
			{Campbell}(2019)}]{deffner2019Book}%
		\BibitemOpen
		\bibfield  {author} {\bibinfo {author} {\bibfnamefont {S.}~\bibnamefont
				{Deffner}}\ and\ \bibinfo {author} {\bibfnamefont {S.}~\bibnamefont
				{Campbell}},\ }\href {\doibase 10.1088/2053-2571/ab21c6} {\emph {\bibinfo
				{title} {Quantum Thermodynamics}}},\ 2053-2571\ (\bibinfo  {publisher}
		{Morgan and Claypool Publishers},\ \bibinfo {year} {2019})\BibitemShut
		{NoStop}%
		\bibitem [{\citenamefont {Ghosh}\ \emph {et~al.}(2019)\citenamefont {Ghosh},
			\citenamefont {Mukherjee}, \citenamefont {Niedenzu},\ and\ \citenamefont
			{Kurizki}}]{Ghosh2019}%
		\BibitemOpen
		\bibfield  {author} {\bibinfo {author} {\bibfnamefont {A.}~\bibnamefont
				{Ghosh}}, \bibinfo {author} {\bibfnamefont {V.}~\bibnamefont {Mukherjee}},
			\bibinfo {author} {\bibfnamefont {W.}~\bibnamefont {Niedenzu}}, \ and\
			\bibinfo {author} {\bibfnamefont {G.}~\bibnamefont {Kurizki}},\ }\href
		{\doibase 10.1140/epjst/e2019-800060-7} {\bibfield  {journal} {\bibinfo
				{journal} {The European Physical Journal Special Topics}\ }\textbf {\bibinfo
				{volume} {227}},\ \bibinfo {pages} {2043} (\bibinfo {year}
			{2019})}\BibitemShut {NoStop}%
		\bibitem [{\citenamefont {Gupt}\ \emph {et~al.}(2021)\citenamefont {Gupt},
			\citenamefont {Bhattacharyya},\ and\ \citenamefont {Ghosh}}]{gupt2021PRE}%
		\BibitemOpen
		\bibfield  {author} {\bibinfo {author} {\bibfnamefont {N.}~\bibnamefont
				{Gupt}}, \bibinfo {author} {\bibfnamefont {S.}~\bibnamefont {Bhattacharyya}},
			\ and\ \bibinfo {author} {\bibfnamefont {A.}~\bibnamefont {Ghosh}},\ }\href
		{\doibase 10.1103/PhysRevE.104.054130} {\bibfield  {journal} {\bibinfo
				{journal} {Phys. Rev. E}\ }\textbf {\bibinfo {volume} {104}},\ \bibinfo
			{pages} {054130} (\bibinfo {year} {2021})}\BibitemShut {NoStop}%
		\bibitem [{\citenamefont {Sinha}\ \emph {et~al.}(2013)\citenamefont {Sinha},
			\citenamefont {Ghosh},\ and\ \citenamefont {Ray}}]{ghosh2013PRE}%
		\BibitemOpen
		\bibfield  {author} {\bibinfo {author} {\bibfnamefont {S.~S.}\ \bibnamefont
				{Sinha}}, \bibinfo {author} {\bibfnamefont {A.}~\bibnamefont {Ghosh}}, \ and\
			\bibinfo {author} {\bibfnamefont {D.~S.}\ \bibnamefont {Ray}},\ }\href
		{\doibase 10.1103/PhysRevE.87.042112} {\bibfield  {journal} {\bibinfo
				{journal} {Phys. Rev. E}\ }\textbf {\bibinfo {volume} {87}},\ \bibinfo
			{pages} {042112} (\bibinfo {year} {2013})}\BibitemShut {NoStop}%
		\bibitem [{\citenamefont {Thierschmann}\ \emph
			{et~al.}(2015{\natexlab{a}})\citenamefont {Thierschmann}, \citenamefont
			{S{\'a}nchez}, \citenamefont {Sothmann}, \citenamefont {Arnold},
			\citenamefont {Heyn}, \citenamefont {Hansen}, \citenamefont {Buhmann},\ and\
			\citenamefont {Molenkamp}}]{thierschmann2015NatureNano}%
		\BibitemOpen
		\bibfield  {author} {\bibinfo {author} {\bibfnamefont {H.}~\bibnamefont
				{Thierschmann}}, \bibinfo {author} {\bibfnamefont {R.}~\bibnamefont
				{S{\'a}nchez}}, \bibinfo {author} {\bibfnamefont {B.}~\bibnamefont
				{Sothmann}}, \bibinfo {author} {\bibfnamefont {F.}~\bibnamefont {Arnold}},
			\bibinfo {author} {\bibfnamefont {C.}~\bibnamefont {Heyn}}, \bibinfo {author}
			{\bibfnamefont {W.}~\bibnamefont {Hansen}}, \bibinfo {author} {\bibfnamefont
				{H.}~\bibnamefont {Buhmann}}, \ and\ \bibinfo {author} {\bibfnamefont
				{L.~W.}\ \bibnamefont {Molenkamp}},\ }\href {\doibase 10.1038/nnano.2015.176}
		{\bibfield  {journal} {\bibinfo  {journal} {Nature Nanotechnology}\ }\textbf
			{\bibinfo {volume} {10}},\ \bibinfo {pages} {854} (\bibinfo {year}
			{2015}{\natexlab{a}})}\BibitemShut {NoStop}%
		\bibitem [{\citenamefont {Hartmann}\ \emph {et~al.}(2015)\citenamefont
			{Hartmann}, \citenamefont {Pfeffer}, \citenamefont {H\"ofling}, \citenamefont
			{Kamp},\ and\ \citenamefont {Worschech}}]{hartmann2015PRL}%
		\BibitemOpen
		\bibfield  {author} {\bibinfo {author} {\bibfnamefont {F.}~\bibnamefont
				{Hartmann}}, \bibinfo {author} {\bibfnamefont {P.}~\bibnamefont {Pfeffer}},
			\bibinfo {author} {\bibfnamefont {S.}~\bibnamefont {H\"ofling}}, \bibinfo
			{author} {\bibfnamefont {M.}~\bibnamefont {Kamp}}, \ and\ \bibinfo {author}
			{\bibfnamefont {L.}~\bibnamefont {Worschech}},\ }\href {\doibase
			10.1103/PhysRevLett.114.146805} {\bibfield  {journal} {\bibinfo  {journal}
				{Phys. Rev. Lett.}\ }\textbf {\bibinfo {volume} {114}},\ \bibinfo {pages}
			{146805} (\bibinfo {year} {2015})}\BibitemShut {NoStop}%
		\bibitem [{\citenamefont {Thierschmann}\ \emph
			{et~al.}(2015{\natexlab{b}})\citenamefont {Thierschmann}, \citenamefont
			{Arnold}, \citenamefont {Mittermüller}, \citenamefont {Maier}, \citenamefont
			{Heyn}, \citenamefont {Hansen}, \citenamefont {Buhmann},\ and\ \citenamefont
			{Molenkamp}}]{thierschmann2015NJP}%
		\BibitemOpen
		\bibfield  {author} {\bibinfo {author} {\bibfnamefont {H.}~\bibnamefont
				{Thierschmann}}, \bibinfo {author} {\bibfnamefont {F.}~\bibnamefont
				{Arnold}}, \bibinfo {author} {\bibfnamefont {M.}~\bibnamefont
				{Mittermüller}}, \bibinfo {author} {\bibfnamefont {L.}~\bibnamefont
				{Maier}}, \bibinfo {author} {\bibfnamefont {C.}~\bibnamefont {Heyn}},
			\bibinfo {author} {\bibfnamefont {W.}~\bibnamefont {Hansen}}, \bibinfo
			{author} {\bibfnamefont {H.}~\bibnamefont {Buhmann}}, \ and\ \bibinfo
			{author} {\bibfnamefont {L.~W.}\ \bibnamefont {Molenkamp}},\ }\href {\doibase
			10.1088/1367-2630/17/11/113003} {\bibfield  {journal} {\bibinfo  {journal}
				{New Journal of Physics}\ }\textbf {\bibinfo {volume} {17}},\ \bibinfo
			{pages} {113003} (\bibinfo {year} {2015}{\natexlab{b}})}\BibitemShut
		{NoStop}%
		\bibitem [{\citenamefont {Pfeffer}\ \emph {et~al.}(2015)\citenamefont
			{Pfeffer}, \citenamefont {Hartmann}, \citenamefont {H\"ofling}, \citenamefont
			{Kamp},\ and\ \citenamefont {Worschech}}]{pfeffer2015PRApp}%
		\BibitemOpen
		\bibfield  {author} {\bibinfo {author} {\bibfnamefont {P.}~\bibnamefont
				{Pfeffer}}, \bibinfo {author} {\bibfnamefont {F.}~\bibnamefont {Hartmann}},
			\bibinfo {author} {\bibfnamefont {S.}~\bibnamefont {H\"ofling}}, \bibinfo
			{author} {\bibfnamefont {M.}~\bibnamefont {Kamp}}, \ and\ \bibinfo {author}
			{\bibfnamefont {L.}~\bibnamefont {Worschech}},\ }\href {\doibase
			10.1103/PhysRevApplied.4.014011} {\bibfield  {journal} {\bibinfo  {journal}
				{Phys. Rev. Applied}\ }\textbf {\bibinfo {volume} {4}},\ \bibinfo {pages}
			{014011} (\bibinfo {year} {2015})}\BibitemShut {NoStop}%
		\bibitem [{\citenamefont {Schaller}\ \emph {et~al.}(2009)\citenamefont
			{Schaller}, \citenamefont {Kie\ss{}lich},\ and\ \citenamefont
			{Brandes}}]{schaller2009PRB}%
		\BibitemOpen
		\bibfield  {author} {\bibinfo {author} {\bibfnamefont {G.}~\bibnamefont
				{Schaller}}, \bibinfo {author} {\bibfnamefont {G.}~\bibnamefont
				{Kie\ss{}lich}}, \ and\ \bibinfo {author} {\bibfnamefont {T.}~\bibnamefont
				{Brandes}},\ }\href {\doibase 10.1103/PhysRevB.80.245107} {\bibfield
			{journal} {\bibinfo  {journal} {Phys. Rev. B}\ }\textbf {\bibinfo {volume}
				{80}},\ \bibinfo {pages} {245107} (\bibinfo {year} {2009})}\BibitemShut
		{NoStop}%
		\bibitem [{\citenamefont {Zedler}\ \emph {et~al.}(2009)\citenamefont {Zedler},
			\citenamefont {Schaller}, \citenamefont {Kiesslich}, \citenamefont {Emary},\
			and\ \citenamefont {Brandes}}]{zedler2009PRB}%
		\BibitemOpen
		\bibfield  {author} {\bibinfo {author} {\bibfnamefont {P.}~\bibnamefont
				{Zedler}}, \bibinfo {author} {\bibfnamefont {G.}~\bibnamefont {Schaller}},
			\bibinfo {author} {\bibfnamefont {G.}~\bibnamefont {Kiesslich}}, \bibinfo
			{author} {\bibfnamefont {C.}~\bibnamefont {Emary}}, \ and\ \bibinfo {author}
			{\bibfnamefont {T.}~\bibnamefont {Brandes}},\ }\href {\doibase
			10.1103/PhysRevB.80.045309} {\bibfield  {journal} {\bibinfo  {journal} {Phys.
					Rev. B}\ }\textbf {\bibinfo {volume} {80}},\ \bibinfo {pages} {045309}
			(\bibinfo {year} {2009})}\BibitemShut {NoStop}%
		\bibitem [{\citenamefont {Wijesekara}\ \emph {et~al.}(2021)\citenamefont
			{Wijesekara}, \citenamefont {Gunapala},\ and\ \citenamefont
			{Premaratne}}]{PRB2021QTT}%
		\BibitemOpen
		\bibfield  {author} {\bibinfo {author} {\bibfnamefont {R.~T.}\ \bibnamefont
				{Wijesekara}}, \bibinfo {author} {\bibfnamefont {S.~D.}\ \bibnamefont
				{Gunapala}}, \ and\ \bibinfo {author} {\bibfnamefont {M.}~\bibnamefont
				{Premaratne}},\ }\href {\doibase 10.1103/PhysRevB.104.045405} {\bibfield
			{journal} {\bibinfo  {journal} {Phys. Rev. B}\ }\textbf {\bibinfo {volume}
				{104}},\ \bibinfo {pages} {045405} (\bibinfo {year} {2021})}\BibitemShut
		{NoStop}%
	\end{thebibliography}
\end{document}